\documentclass[twocolumn, trackchanges]{aastex631} 

\usepackage{graphicx} 
\usepackage{float}
\usepackage{hyperref}
\usepackage{amssymb}
\usepackage{amsmath}
\usepackage{breqn}
\usepackage{longtable}
\usepackage{booktabs}

\usepackage{epsf}
\usepackage{listings}
\usepackage{rotating}
\usepackage{capt-of}

\newcommand\GalCEM{{\tt GalCEM}}
\newcommand{\Msun}{$M_{\odot}$}
\newcommand{\deriv}{\mathrm{d}}

\newcommand{\myfig}{{Fig. }}
\newcommand{\myeq}{{Eq. }}

\received{\today}
\revised{}
\accepted{}
\submitjournal{ApJS}

\shorttitle{\emph{GalCEM} method presentation}
\shortauthors{Gjergo et al.}
\graphicspath{{./}{figures/}}

\begin{document}

\title{\GalCEM\, I -- An Open-Source Detailed Isotopic Chemical Evolution Code
\footnote{Released on MM, DDth, YYYY}
}

\author[0000-0002-7440-1080]{Eda Gjergo}
\affiliation{School of Physics and Technology, Wuhan University, Wuhan 430072, China.}\email{eda.gjergo@gmail.com}
\affiliation{School of Astronomy and Space Science, Nanjing University, Nanjing 210093, People's Republic of China}
\affiliation{Key Laboratory of Modern Astronomy and Astrophysics (Nanjing University), \\Ministry of Education, Nanjing 210093, People's Republic of China}

\author[0000-0003-1004-4632]{Aleksei G. Sorokin}
\affiliation{Department of Applied Mathematics, Illinois Institute of Technology, RE 220, 10 W. $32^{\textrm{\footnotesize nd}}$ St., Chicago IL 60616, USA}

\author[0000-0002-2670-5709]{Anthony Ruth}
\affiliation{Cubic PV, 1807 Ross Ave, Ste 333 Dallas, Texas 75201, USA}

\author[0000-0001-9715-5727]{Emanuele Spitoni}
\affiliation{Universit\'{e} C\^{o}te d’Azur, Observatoire de la C\^{o}te d’Azur, CNRS, Laboratoire Lagrange, \\Bd de l’Observatoire, CS 34229, 06304 Nice cedex 4, France}

\author[0000-0001-7067-2302]{Francesca Matteucci}
\affiliation{Dipartimento di Fisica, Sezione di Astronomia, Universit\`{a} di Trieste, Via G. B. Tiepolo 11, I-34143 Trieste, Italy}
\affiliation{INAF, Osservatorio Astronomico di Trieste, Via Tiepolo 11, I-34143 Trieste, Italy}
\affiliation{INFN, Sezione di Trieste, Via A. Valerio 2, I-34127 Trieste, Italy}
\affiliation{IFPU—Institute for the Fundamental Physics of the Universe, Via Beirut, 2, I-34151 Trieste, Italy}

\author[0000-0002-8174-0128]{Xilong Fan}
\affiliation{School of Physics and Technology, Wuhan University, Wuhan 430072, China.}

\author[0000-0001-8405-2921]{Jinning Liang}
\affiliation{School of Physics and Technology, Wuhan University, Wuhan 430072, China.}

\author[0000-0003-0636-7834]{Marco Limongi}
\affiliation{Istituto Nazionale di Astrofisica - Osservatorio Astronomico di Roma, Via Frascati 33, I-00040, Monteporzio Catone, Italy}
\affiliation{Kavli Institute for the Physics and Mathematics of the Universe, Todai Institutes for Advanced Study, the University of Tokyo, Kashiwa, Japan 277-8583 (Kavli IPMU, WPI)}

\author[0000-0003-3783-3973]{Yuta Yamazaki}
\affiliation{National Astronomical Observatory of Japan, 2-21-1 Osawa, Mitaka, Tokyo 181-8588, Japan}
\affiliation{Graduate School of Science, The University of Tokyo, Hongo, Bunkyo-ku, Tokyo 11-0033, Japan}

\author[0000-0003-3083-6565]{Motohiko Kusakabe}
\affiliation{School of Physics, and International Research Center for Big-Bang Cosmology and Element Genesis, Beihang University, Beijing 100083, P. R. China}
\affiliation{National Astronomical Observatory of Japan, 2-21-1 Osawa, Mitaka, Tokyo 181-8588, Japan}

\author[0000-0002-8619-359X]{Toshitaka Kajino}
\affiliation{School of Physics, and International Research Center for Big-Bang Cosmology and Element Genesis, Beihang University, Beijing 100083, P. R. China}
\affiliation{National Astronomical Observatory of Japan, 2-21-1 Osawa, Mitaka, Tokyo 181-8588, Japan}
\affiliation{Graduate School of Science, The University of Tokyo, Hongo, Bunkyo-ku, Tokyo 11-0033, Japan}



\begin{abstract}
This is the first of a series of papers that will introduce a user-friendly, detailed, and modular {\tt GALactic Chemical Evolution Model}, \GalCEM, that tracks isotope masses as a function of time in a given galaxy. The list of tracked isotopes automatically adapts to the complete set provided by the input yields. 
The present iteration of \GalCEM\, tracks 86 elements broken down in 451 isotopes. The prescription includes massive stars, low-to-intermediate mass stars, and Type Ia supernovae as enrichment channels. We have developed a preprocessing tool that extracts multi-dimensional interpolation curves from the input yield tables. These interpolation curves improve the computation speeds of the full convolution integrals, which are computed for each isotope and for each enrichment channel.  We map the integrand quantities onto consistent array grids in order to perform the numerical integration at each time step. The differential equation is solved with a fourth-order Runge-Kutta method.

We constrain our analysis to the evolution of all the light and intermediate elements from carbon to zinc, and lithium. Our results are consistent up to the extremely metal poor regime with Galactic abundances. We provide tools to track the mass rate change of individual isotopes on a typical spiral galaxy with a final baryonic mass of $5\times 10^{10}$ \Msun. Future iterations of the work will extend to the full periodic table by including the enrichment from neutron-capture channels as well as  spatially-dependent treatments of galaxy properties. \GalCEM\, is publicly available at \url{https://github.com/egjergo/GalCEM}.

\end{abstract}


\section{Introduction}

 \begin{figure*}[!htpb]
\centering \includegraphics[width = \textwidth]
{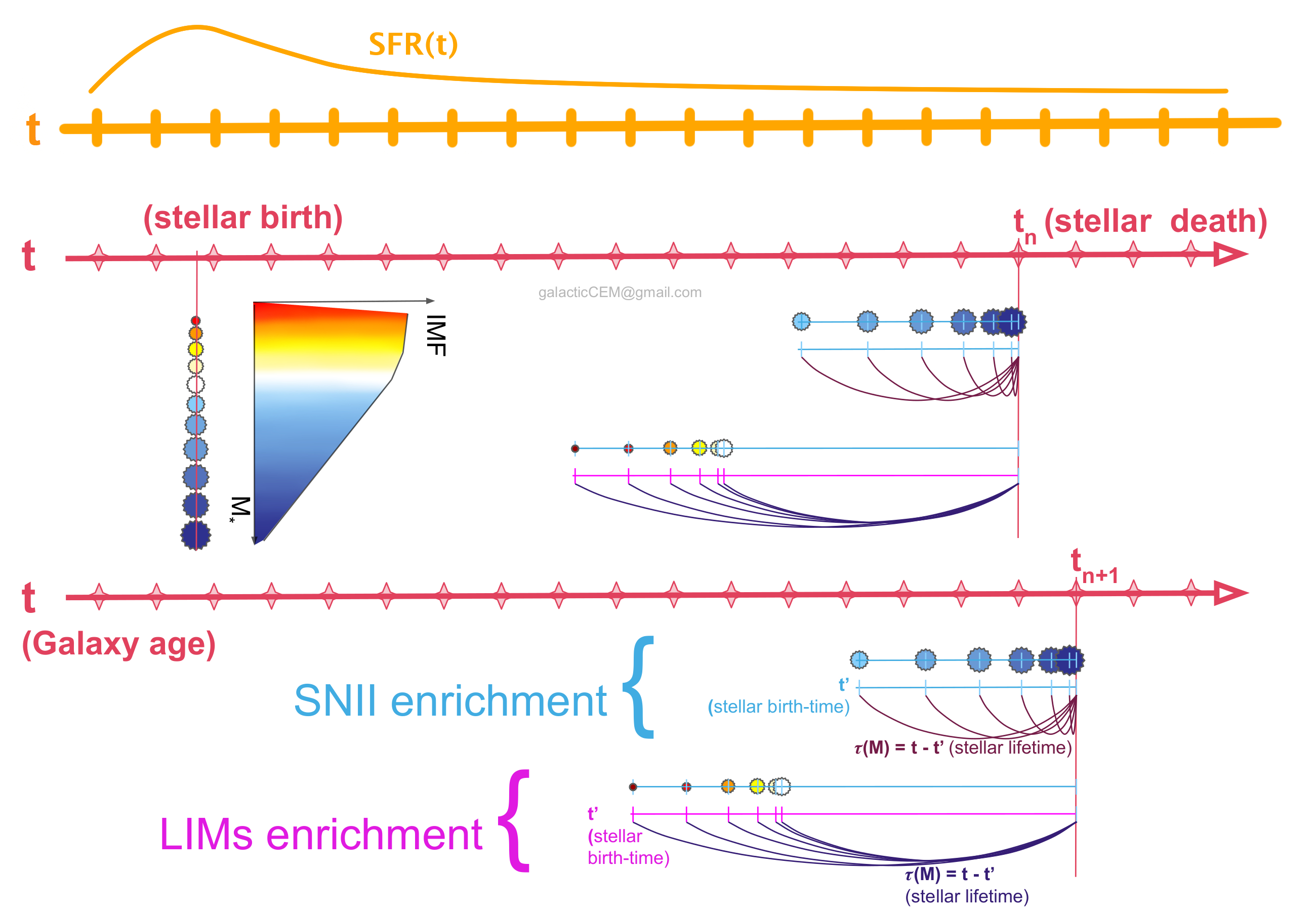}
\caption{Diagram of the GCE rationale. At each time step, the SFR, $\psi(t)$, determines how much total gas mass is converted into stellar mass. The distribution of stellar masses $M_*$ follows the given IMF, $\phi(M_*)$. At each subsequent time step, the GCE convolution integral determines how many stars die at $t_n$, and therefore how much enrichment these stars return to the gas mass. The stellar birth-time $t'$, the baby blue and magenta grids, and lifetimes $\tau(M_*,Z_*)$, i.e. the curves connecting birth-time with $t_n$, are updated at each time step. Each enrichment channel will cover a mass/lifetime range specific to its process (SNCC -- or equivalently for our purposes, SNII -- and LIMs are included in the graphic). Onto these enrichment channel grids we compute the integrals, as explained in \myfig \ref{fig:integrGrid}.}\label{fig:diagram}
\end{figure*}

Galactic Chemical Evolution (GCE) is the field that investigates how the chemical makeup of a galaxy changes with time. To do that, GCE must first consider the astrophysical events where each isotope may be synthesized. These events are point-like sources of enrichment, which occur with varying rates throughout the galaxy and across time. To model how such events are distributed in a variety of galactic morphologies, the field of GCE has developed rate equations which track the abundance of chemical species.
A galactic morphology is reconstructed by means of stellar birthrate functions. Stellar lifetimes will then determine when the chemical enrichment will contaminate the galactic media.
Hereafter we will refer to``astrophysical events" and``enrichment channels" interchangeably \citep[for a recent review, see][]{matteucci21}.

The theory of GCE dates its origins to the mid 1950's with the pioneering works of \citet{salpeter55, salpeter59}, and \citet{schmidt59}, who laid the foundations to the formalism best explained in \citet{schmidt63} and subsequently in \citet{tinsley80}. A schematic representation of the formalism can be seen in \myfig \ref{fig:diagram} as it applies to one-zone models \citep[e.g.,][]{talbot71, timmes95}, i.e., the treatment of galaxies as homogeneous environments where the gas is mixed instantaneously. 

In order to formulate the luminosity function evolution of main-sequence stars, \citet{salpeter55} proposed the existence of an ``original mass function", now called initial mass function (IMF) -- or, the  distribution, with respect to stellar mass $M_*$, of stars at their birth in a single stellar population. This IMF distribution is color-coded in \myfig \ref{fig:diagram} according to main-sequence colors. The IMF is one of the two components of the birthrate function $\mathcal{B}(M_*,t)$, the other being the star formation rate \citep[SFR, ][]{salpeter59, schmidt59}. The ansatz is that the birthrate function is separable with respect to stellar mass and time and is represented by the product of SFR and IMF. This is a simplistic ansatz, already challenged by numerous works \citep{yan17,yan21}, but it nonetheless proved to be valuable in GCE. 

We must also stress the difference between backward and forward approaches. Cosmological simulations generally follow a forward approach, wherein stellar birth is tracked first, then the chemical enrichment is projected onto future time steps. A classic GCE approach, instead, follows a backward approach, namely it reconstructs past stellar distributions by tracking stellar death rates at every time-step \citep{matteucci12}.

\begin{table*}[!htbp]
   \renewcommand{\arraystretch}{1.2}
   \centering
   \begin{tabular}{@{} l |  l @{}} 
      \hline
       Symbol    & Description \\
      \hline \hline
      $i$                                          &   index of the tracked isotope (the chemical species)  \\ 
      $k$                                          &   mass grid integration index  \\ 
      $n$                                          &   time step index  \\ 
      $P$                                         & the label that identifies an astrophysical process \\ 
      & that is a channel of chemical enrichment \\\hline
      $t_G$                                     &   final age of the galaxy \\
       $t$                                         &  age vector of the galaxy \\
      $t'(M_*,Z_*)$                                 & birth time of a star of mass $M_*$ \\
      $\tau(M_*,Z_*) = t - t'(M_*,Z_*)$          & lifetime of a star of mass $M_*$ and metallicity $Z_*$ \\ \hline
      $M_{*}$                                  & mass of a single star \\ 
      $M_{\rm P,l}, M_{\rm P,u}$                 & lower and upper mass limits for stars in the integrals \\
      & (mass limits are process-specific) \\
      $M_{\rm inf}(t)$                            &  baryonic mass of the galaxy as a function of time \\
      & (determined by the infall rate)\\
      $M_{\rm gas}(t)$                          & gas mass of the galaxy as a function of time \\
      $M_{\rm i,gas}(t)$                        & gas mass of the isotope $i$ in the galaxy as a function of time \\
                                                     & s.t. $\sum_i M_{\rm i,gas}(t) = M_{\rm gas}(t)$   \\
      $M_{\rm *,tot}(t)$                          & star mass of the galaxy as a function of time \\
                                                     & $M_{\rm inf}(t) = M_{\rm gas}(t) + M_{\rm *,tot}(t)$   \\ \hline
      $\nu$                                                          & star formation efficiency \\
      $\psi(t)  $                                                   & star formation rate (SFR) equivalent to $\dot{M}_{\rm *,tot}(t)$  \\
      $\phi(M_*)$                                                   & initial mass function (IMF) \\
      $Z(t)$                           & the metallicity, or the mass fraction of all of the chemical \\
                                           & elements with the exclusion of H and He \\  
      $Y_P(i, M_*, Z_*)$                 & Yields: tabulations of the mass fractions $M_i/M_{\rm *,R}$,  \\
      					       &  where $M_{\rm *,R}$ is the total mass returned  to  \\
					       &  the interstellar medium by a star of mass $M_*$ \\
					       & and of initial metallicity $Z_*$, for the astrophysical process $P$\\ \hline
   \end{tabular}
   \caption{Glossary of GCE Symbols and Quantities of Interest, as They Have Been Used throughout the Present Article.}
   \label{tab:components}
\end{table*}

The interstellar medium (ISM) will generally be enriched by the nucleosynthesis products (yields) generated in individual astrophysical sites -- be them supernovae (SNe), asymptotic giant branch (AGB) winds, or coalescence events, to name a few. GCE must integrate the occurrence of such events over the lifetime of the galaxy. To do that, GCE must couple the birthrate function with stellar lifetimes to extrapolate a ``death-rate function" and hence learn when the enrichment events will occur. These, coupled with the yields, define how the abundance of each isotope evolves with time. In \myfig \ref{fig:diagram}, at each time step $t_n$, GCE models reconstruct the birth time and properties of all stars that die within that time step and compute their integrated contribution to the enrichment from each astrophysical site. A first exhaustive review connecting nuclear physics to astrophysical sites undergoing nucleosynthesis is the cornerstone work by \citet{burbidge57}, famously referred to as B2FH. 

Over the decades, chemical evolution has provided useful insights on galaxy properties. For example, \citet{eggen62} first discovered that halo stars are old and metal poor while disk stars are young and metal-rich -- hence leading to the understanding that disk stars formed along with gas infall -- and that such stars offer a snapshot to the chemical composition of their parent star-forming cloud.

Based on the cornerstone work by \citet{salpeter55}, 
\citet{salpeter59} and \citet{schmidt59} independently developed a very similar GCE formalism that laid the foundation for all subsequent works \citep[reviews throughout the years include][]{tinsley80, prantzos08, pagel09, matteucci12, matteucci21}. In particular,  \citet{talbot71} developed a numerical solution to the full convolution equations in a work that did not require the metallicity to grow monotonically. \citet{pagel75} developed a modified GCE model based on the simple model which included prompt initial enrichment, the early version of metal-enhanced star formation, and inhomogeneous collapse and infall. \citet{chiosi80} proposed an open model with gas infall and outflow. \citet{chiosi80m} and later also \citet{portinari00} developed a radially-dependent disk GCE model, which was later improved and expanded in \citet{spitoni11} where they found that an inside-out formation scenario (and/or a variable flow) in spiral disks is necessary to reproduce Galactic abundance gradients. The effect of stellar migration in the Milky way was investigated by \citet{schoenrich09, spitoni15} and \citet{johnson21}.
Gas radial flow as well as stellar mixing were considered in a multi-zone GCE model in \citet{chen22}. A Bayesian approach was undertaken in \citet{cote17, rybizki17} and \citet{spitoni20}, this last work in particular applied Markov chain Monte Carlo methods to a two-infall Milky Way formation scenario.

\citet{spitoni17} proposed new analytical solutions to a GCE model that describes SDSS galaxies only as a function of infall timescales, infall masses, and mass loading factors. 

A handful of GCE models have also been released as public codes. \citet{andrews17} presented flexCE, a flexible one-zone chemical evolution code which was recently used in \citet{hasselquist21}.  \citet{cote17} presented the One-zone Model for the Evolution of GAlaxies (OMEGA) code within the NuGrid Python Chemical Evolution Environment (NuPyCEE). OMEGA is paired with the Stellar Yields for Galactic Modeling Applications \citep[SYGMA,][]{ritter18} which computes the ejecta of single stellar populations (SSP) to reconstruct the enrichment of a galaxy. \citet{yan17} proposed a one-zone model (GalIMF) in which they adopted a variable integrated galactic initial mass function (IGIMF). \citet{rybizki17} developed Chempy, which parametrizes open one-zone models within a Bayesian framework.
And more recently, \citet{johnson20} developed the Versatile Integrator for Chemical Evolution (VICE), an efficient and user-friendly code that shortens computation times by applying IMF-averaged yields to the SN enrichment by core collapse SNe.

There has been a fortunate surge in high precision surveys and instruments — some concluded very recently, some ongoing still — that are providing unprecedented levels of precision to constrain GCE. These include LAMOST \citep[Large Sky Area Multi-Object Fiber Spectroscopic Telescope,][]{cui12, deng12, zhao12}, Gaia-ESO \citep{gilmore12, randich22}, Gaia DR3 \citep{recio-blanco22} whose $\alpha$-element abundances have already been investigated with the two-infall model \citep{spitoni22},  RAVE \citep{kordopatis13, steinmetz20}, APOGEE \citep{majewski17, ahumada20, abdurro'uf22} with spectra in the IR, SDSS with photometric optical-nearIR bands, GALAH \citep{deSilva15, buder21} with the goal of obtaining abundances for 30 elements within the Galaxy, Vista Variables in the Via Lactea \citep[VVV][]{minniti10}  with a focus on the Galactic Bulge, MUSE \citep{bacon10} with a highly resolved integrated field spectroscopy unit and a broader scope out to high redshifts, 4MOST \citep{dejong19,walcher19} which will measure spectroscopic redshifts of X-ray-identified groups and galaxy clusters, and other upcoming instruments such as JWST \citep{gardner06}. Notably, the Gaia-ESO mission was devised with the purpose of precisely mapping the 3D locations and motions of billions of stars within our galaxy; the low-resolution LAMOST spectrograph has been able to collect simultaneously up to 4000 spectra over huge volumes of the Galaxy, with both a large aperture and a large field of view. The analysis of these data which in most cases is coupled with stellar dynamics make it so that this 70-year-old field is mature enough to properly constrain multi-zone statistical analysis and provide further information on the evolutionary history of the galaxy. { Before investigating such pursuits, we lay our foundations by reaching a benchmark with previous studies by means of one-zone modeling.}

{ With \GalCEM\, we offer a public code that adapts to the complete nuclide list of the chosen yields. \GalCEM\, can solve the GCE integrodifferential equation including infall, outflow, SFR and fully convolved returned ejecta across the whole main-sequence. By default, low-mass stars, massive stars, and Type Ia supernova yields are always included. In this work we limit our analysis to these three enrichment channels only. }

The article is structured as follows: in Section \ref{sec:method} we introduce the GCE formalism and we present our numerical solution to our adopted general GCE equations. In Section \ref{sec:results} present some preliminary results, as well as the data products available at this stage in \GalCEM, and we evaluate aspects of the code performance. Finally in Section \ref{sec:discussion} we summarize the article and we discuss future aims and scientific goals we expect to achieve with this tool.

\begin{figure*}[htbp]
\centering \includegraphics[width =\textwidth]
{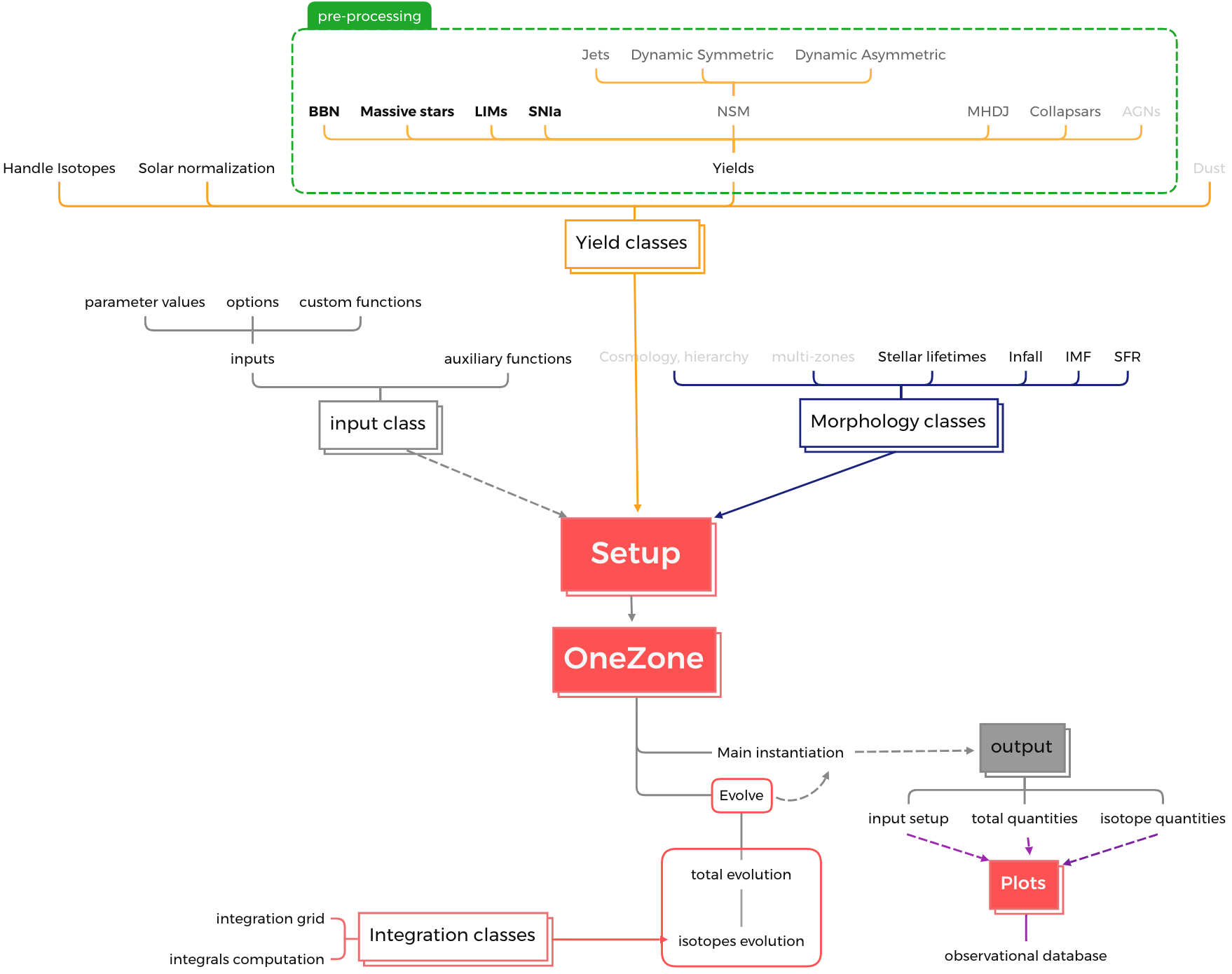}
\caption{\GalCEM\, flowchart. A description of the figure is provided in Section \ref{sec:workflow}.} \label{fig:flowchart}
\end{figure*} 

\section{Method} \label{sec:method}

In this section we present both the theoretical framework on which we base our computation and the numerical solution we developed. At present, \GalCEM\, is available as a one-zone code, i.e., we adopt an instantaneous mixing approximation  and the whole galaxy is treated as a single homogeneous zone. The solution has been implemented in the Python code \GalCEM, publicly available on GitHub\footnote{\url{https://github.com/egjergo/GalCEM/galcem}}. 

We follow the conventional formalism, so the general GCE equation can be succinctly expressed as \citep{pagel09}:
\begin{equation}\label{eq:GCEshort}
\dot{M}_{\rm i,gas}(t) = \mathcal{I}(t) - X_{\rm i,gas}\psi(t) + \sum_P{R}_{\rm P,i}(t) - \mathcal{O}(t),
\end{equation} 
where $\dot{M}_{\rm i,gas}(t)$ is the mass rate of change of the gas-phase component of a chemical species or isotope $i$, {$\mathcal{I}(t)$ represents the infall term} while $\mathcal{O}(t)$ is the outflow term. $X_{\rm i,gas}\psi(t)$ is the mass fraction of the $i$-th isotope ($X_{\rm i,gas} = M_{\rm i,gas}/M_{\rm tot,gas}$) that is subtracted from the total gas component due to star formation. $\psi(t)$ refers to the SFR, i.e., of the total gas mass that forms stars at a given time step. ${R}_{\rm P,i}(t)$ is the rate of the $i$-th component returned  to the gas phase by each enrichment channel $P$. At present, such enrichment channels include the final stages of low-to-intermediate mass stars (LIMs) and their enrichment of the ISM through asymptotic giant branch (AGB) winds, the death of massive stars as core collapse supernovae (SNCC), and Type Ia supernovae (SNIa).
In Table \ref{tab:components} is the synopsis all of the main symbols involved in the GCE formalism. The symbols which have not been introduced in the present paragraph will be explained in due course within this section\footnote{The present article follows the common square bracket notation of the logarithmic abundance: for an element or isotope $A$, the abundance relative to another element $B$ will be:
\begin{align}
\begin{split}
\mathrm{[A/B]} =& \log\left(M_{\mathrm{A}}/M_{\mathrm{B}}\right) - \log\left(M_{\mathrm{A}}/M_{\mathrm{B}}\right)_{\odot}\\
 =&\log\left(\mu_A X_{\mathrm{A}}/\mu_{\mathrm{B}}X_{\mathrm{B}}\right) - \log\left(\mu_A X_{\mathrm{A,\odot}}/\mu_{\mathrm{B}}X_{\mathrm{B,\odot}}\right)\\ 
=& \log\left(X_{\mathrm{A}}/X_{\mathrm{B}}\right) - \log\left(X_{\mathrm{A}}/X_{\mathrm{B}}\right)_{\odot},
\end{split}
\end{align}
where $\mu$ is the atomic weight of a chemical species and $\odot$ marks solar values.}.

The galaxy begins by having no mass. The only source of mass growth is provided by the infall term that accretes gas from a primordial (or otherwise very metal poor) intergalactic medium\footnote{Here we follow the gas-phase definition of metallicity ${Z(t)} = {\left(M_Z(t)=M_{\rm gas}(t) - M_{\textrm{H}}(t) - M_{\textrm{He}}(t)\right)}/{M_{\rm gas}(t)}.$, where ``metals" are all elements aside from H and He}. 
We conform our solar metallicity to the value derived in \citet{asplund09} of  $Z_{\odot} = 0.0134$. An environment, star, or system is defined as very metal poor whenever $< 10^{-2} Z_{\odot}$, and extremely metal poor for $< 10^{-3} Z_{\odot}$ \citep[e.g., ][]{nomoto13}.

The presence of gas triggers star formation, described by the SFR. We take the SFR to be a function of the total galaxy mass and the total gas mass at any given time. We assume that the mass distribution of stars is described at every time step by the IMF. The mass of newly synthesized isotopes at any time will be given by the convolution of the SFR with the IMF and yields, as described by \myeq \ref{eq:general}, which is the full integrodifferential version of \myeq \ref{eq:GCEshort} and the equation being solved by \GalCEM. The majority of the calibrations in this section are taken from \citet{molero21a,molero21b}. In the rest of this Methods section we will provide detailed information about every necessary GCE ingredient.

\subsection{General workflow in \GalCEM }\label{sec:workflow}
 \myfig \ref{fig:flowchart} represents the \GalCEM\, flowchart. Its design follows a simple principle, namely that GCE considers individual nucleosynthetic enrichment channels and integrates the occurrence of each event across both time and across the distribution of such events in a galaxy, as constructed through the combination of birth-rate functions and stellar lifetimes. Each event associated with each enrichment channel is represented by individual points for each isotope in yield tabulations. These point sources of enrichment are treated in the yield class. Birth rate and stellar lifetimes are instead treated in the morphology class. While \GalCEM\, follows well-established literature prescriptions on the GCE theory, the design and numerical solution to the integrodifferential equation is original to this work.
 
 \GalCEM\, contains a preprocessing yield tool, and requires the input from three classes at setup: inputs.py, yields.py, and morphology.py. Regarding the preprocessing yield tool, it extracts the interpolation surfaces for each yield tabulation selected for a run. 
  
  {\tt morphology.py} contains the ingredients that characterize the galaxy properties (namely, IMF, SFR, infall, and stellar lifetimes). yields.py imports the necessary yield properties, including the preprocessed interpolations, and provides the tools to combine them within the setup class in Main. 
  
  {\tt yields.py} also contains the solar normalization class, and the class that extracts a combined list of unique isotopes from all of the selected yields. The OneZone class in Main opens and writes the outputs and runs the {\tt evolve} function, which computes the time-step-dependent GCE quantities. The global quantities (namely, total gas and total stellar mass) are computed by the {\tt total\_evolution} function, while the isotope-dependent quantities are computed by {\tt isotopes\_evolution}. { Light-gray font colors represent \GalCEM\, features that will be presented in the near future}

 \GalCEM\, runs on {\tt Python3}. After importing the package, an inputs object must be defined. The inputs object is customizable\footnote{The full list of input parameters is in \url{https://github.com/egjergo/GalCEM/blob/main/galcem/classes/inputs.py}}.
 A minimum working example has the following form\footnote{The minimum working example script is located in \url{https://github.com/egjergo/GalCEM/blob/main/examples/mwe.py}}:
 \begin{lstlisting}[language=Python]
import galcem as gc
inputs = gc.Inputs()
oz = gc.OneZone(inputs,outdir=MYDIR)
oz.main()
 \end{lstlisting}
By defining the {\tt oz} object, the user initializes the setup, including a series of properties like the time vector, the total gas mass, and the complete list of isotopes generated from the chosen yields. The actual run is launched by calling the {\tt main} function onto the {\tt oz} object.
  
The parameters of the simulation can either be set within the {\tt inputs.py} parameters or they can be edited manually in a script before executing a run. For example, the size of the time step, in Gyr, can be easily changed with the following:
{ \begin{lstlisting}
 inputs.ntime step = .2
 \end{lstlisting}}

\GalCEM\, has been published as a  Python Package Index Project, and for {\tt pip} users can be installed with the terminal or conda command:
\begin{lstlisting}
pip install galcem
\end{lstlisting}

Alternatively, the interested user who prefers to work on the cloud can request a JupyterHub account.\footnote{Available at  \url{https://galcem.space/}.}

\subsection{Infall rate and Outflows}\label{sec:infall}
Infall { as well as outflow} rates play an essential role in GCE models, in that they can characterize the dynamics of galaxy formation \citep[e.g., see ][]{pagel09}. { Concerning infall rates,} declining models are favored because they predict metallicity distributions more closely consistent with those of G-dwarf Milky Way disk stars \citep{larson74, larson76, matteucci96}. 

\citet{spitoni21} recently investigated the assumption of a declining infall, as well as the impact of inflow and outflow on the cumulative metallicity distribution of galaxies as a function of their stellar mass.
In this work we therefore consider a simple single exponential infall rate described by:
\begin{equation} \label{eq:infall}
\mathcal{I}(t)=\dot{M}_{\rm inf}(t) = I_0 e^{-t/\tau_{inf}},
\end{equation}
where $\dot{M}_{\rm inf}(t) = \deriv {M}_{\rm inf}(t)/ \deriv t$ is the rate of gas mass from the extragalactic medium falling into the gravitational potential of a galaxy with time.  We assume that only gas falls into a galaxy's potential. In this preliminary work, we ignore the outflow term in \myeq \ref{eq:GCEshort}. This simplification is justified by the results of \citet{spitoni09}, where they found that the outflow timescales as inferred from the orbit times of clouds subject to the Galactic gravitational potential should be of the order of 0.1 Gyr. The impact of such outflow timescales on GCE models is negligible.

The infall timescale {$\tau_{inf}$} varies depending on the formation history of a galaxy. { Shorter timescales imply more rapid star formation histories and are associated with elliptical galaxies \citep[][]{pipino04}, while more relaxed timescales are associated with spiral or dwarf galaxies. In \citet[][]{molero21a}, the timescale that reproduces the Milky Way thin disk is fine-tuned to 7 Gyr, while 0.2 Gyr is suitable for elliptical galaxies \citep[or 0.05 Gyr for some dwarf galaxies in ][]{molero21b}. }
$I_0$ has units of [\Msun\,/ Gyr], with \Msun\,being the solar mass. It is normalized so that:
\begin{equation}
\int_{0}^{t_G}\left(\dot{M}_{\rm inf}(t) - \dot{M}_{\rm out}(t)\right)\textrm{d}t = M_{\rm tot},
\end{equation}
where $t_G$ and $M_{\rm tot}$ (e.g., $M_{\rm inf}(t_G) = M_{\rm tot}$) represent the present-day age and total baryonic mass of the galaxy, respectively. Given that we do not treat dynamics, gravitational components including dark matter are beyond the scope of this work.

Similarly, { in the specific case of no outflow,} the total baryonic mass of a galaxy as a function of time is given by:
\begin{equation}
M_{\rm inf}(t) =  M_{\rm tot}(t) = \int_{0}^{t}\dot{M}_{\rm inf}(t) \textrm{d}t.
\end{equation}

\paragraph{Infall rate in \GalCEM}
\,\GalCEM\, computes the infall mass $M_{\rm inf}(t)$ as well as the total mass $M_{\rm tot}(t)$ at the setup stage when the one-zone object {\tt oz} is defined. We take the final total baryonic mass to be $M_{\rm tot}(t_G)=5\times10^{10}$ \Msun, with $t_G=13.7$ Gyr being the present age of the Galaxy, and the infall timescale $\tau_{inf}=7$ Gyr.

\paragraph{Outflows}
{ 
Galactic outflows are an integral component galaxy evolution modeling. They have been invoked since the late 70s \citep[e.g.,][]{deyoung78} as the source of metal-enrichment in the intracluster medium. Recent Chandra X-ray observations of M82 outflows suggest that the hot halo gas is being mass-loaded with cold-phase material \citep{lopez20}. This material can be very metal-rich, like in the case of the edge-on star-forming galaxy Mrk 1485, whose outflows are about 1.6 times the ISM metallicity \citep{cameron21}. How these outflows impact the chemical evolution of a galaxy is not yet a settled issue. A number of papers which do incorporate outflows find that it significantly impacts the evolution of chemical abundances \citep[e.g.,][]{andrews17, weinberg17, trueman22}.

However, studies on Galactic fountains triggered by core-collapse SNae associated with the solar annulus  \citet[][]{melioli08, spitoni08} find that that outflow ejecta should fall back into the close proximity to the same Galactocentric regions. Furthermore, the typical fall-back delay of such clouds is found to be $\sim$ 0.1 Gyr \citep{spitoni09}. Despite the dynamical complexity of such phenomena, their impact on the chemical evolution appears to be negligible. In fact, a number of GCE models which assume no outflow are able to reproduce observations \citep[e.g., ][]{minchev13, romano19}. But the issue is further complicated by another consideration: if the gas is ejected onto the circumgalactic medium, semi-analytic models find that this gas will become available again for cooling on much longer timescales, of the order of several Gyr \citep[][]{faerman22}. 

\paragraph{Outflows in \GalCEM}
To set a benchmark for comparison with other works, in this first paper we set the outflow to zero, but the user can easily edit {\tt inputs.wind\_efficiency} to implement a dimensionless parameter $0< \omega < 1$ so that the outflow may be proportional to the SFR, $\mathcal{O}(t) = \omega \psi(t)$. The proportionality of the outflow to the SFR follows \citet[][]{bradamante98}, where the galactic winds originate whenever the thermal energy of the gas exceeds its binding energy. Alternatively, the user can comment out the {\tt self.wind\_efficiency = 0} override in {\tt classes/inputs.py} to restore the default parameter for the given morphology.
}

\subsection{Initial Mass Function (IMF)}

The initial mass function, IMF, is the number distribution of stars in a galaxy as a function of stellar mass. Its original single-power-law formulation dates back to \citet[][]{salpeter55}:
\begin{equation}\label{eq:IMF}
\phi(M_*) =\mathrm{d} N_*/\mathrm{d} M_*=     \phi_0 (M_*/M_{\rm \odot})^{-x},
\end{equation}
where $M_*$ is the stellar mass of individual stars, normalized to the solar mass \Msun, and the best fit for the power law index was found to be $x=2.35$. The IMF is one component of the birthrate function, the other being the SFR. Given that we already express the SFR in units of solar masses over time, we normalize the mass-weighted IMF to 1: $\int^{M_u}_{\rm M_l} M_* \phi (M_*) \mathrm{d}M_* = 1$. 

{ A more appropriate IMF which is representative of the stellar distribution in the Milky Way is the universal IMF \citep[][]{kroupa01}:
\begin{equation}\label{eq:canonicalIMF}
    \phi (M_*) =\left\{ \begin{array}{ll}
    2\phi_0 M_*^{-\alpha_1}, \hspace{0.65cm} M_l\leq M_*/M_{\rm \odot}<0.50 \,, \\
    \phi_0 M_*^{-\alpha_2}, \hspace{0.65cm} 0.50\leq M_*/M_{\rm \odot}<1.00 \,, \\
    \phi_0 M_*^{-\alpha_3}, \hspace{0.65cm} 1.00\leq M_*/M_{\rm \odot}< M_u \,.
    \end{array} \right.
\end{equation}
This IMF is now commonly referred to as canonical IMF, and it is a piecewise power law that accounts for the fact that the low-mass end of the stellar mass distribution is not as steep as the high-mass end. For a historical overview of the field, see \citet[][]{kroupa13}.}

\paragraph{IMF in \GalCEM}
{ For the present paper we adopted the \citet[][]{kroupa01} canonical IMF, with $\alpha_1=1.3$ and $\alpha_2=\alpha_3=2.3$, according to the canonical values. In \myeq \ref{eq:canonicalIMF} we left the break at $M_*/$\Msun$>1$ because it is representative of \GalCEM's implementation, i.e., the user may readily explore top-light or top-heavy IMFs. We chose  $M_l = 0.08$ to $M_u = 120$ \Msun to be the span of the mass limits for the normalization, in accordance with the upper mass limit of our chosen SNCC yields \citep[][]{limongi18}. The Salpeter IMF is also an available option ({\tt inputs.IMF\_option = 'Salpeter55'}).} \GalCEM\, can take custom IMF laws, as long as they are only dependent on the stellar mass.  Variable IMFs that depend e.g. on the metallicity and SFR, such as the integrated galaxy-wide IMF \citep[IGIMF][]{yan17, jerabkova18, yan21}, will be investigated in the near future.

\subsection{The Star Formation Rate (SFR)}\label{sec:SFR}

One of the prevalent SFR prescriptions, available by default in \GalCEM, is:

\begin{equation}\label{eq:SFR}
\psi(t) = \dot{M}_{\rm *,tot}(t)  = \nu \left(\frac{M_{\rm gas}(t)}{M_{\rm tot}(t)}\right)^{\kappa},
\end{equation}

is a Kennicutt-Schmidt definition \citep{kennicutt98} of the SFR, where $\dot{M}_{\rm *,tot}(t)$ is the rate of total stellar mass formed as a function of time, $M_{\rm gas}(t)$ is the gas mass at time $t$, and $M_{\rm tot}(t)$ is the total baryonic mass fallen within the galaxy by time $t$. $M_{\rm tot}(t)$ is initially  added to the gas component. This gas mass will be partly composed of the pristine infall gas mass, but it will also be enriched over time by the returned ejecta of the astrophysical sites $P$.  This returned mass is computed by the convolution integrals shown in \myeq \ref{eq:birth timerate}.

The star formation efficiency (SFE, $\nu$) has units of [Gyr$^{-1}$] for a SFR defined in terms of surface densities, while {$\kappa$} is a dimensionless parameter that varies depending on the morphology. $\kappa$ normally ranges between 1 and 2. { The best fit found in \citet[][]{kennicutt98} is 1.4}. 

{ This single SFR power law is however a simplification. \citet{kennicutt21} find  breaks in both the power-law index and zero-points of the star formation law between starbursting and non-starbursting galaxies. \citet{ellison21} demonstrate that there is significant galaxy-to-galaxy variations in this trend, suggesting that a single relation is not universally applicable. Much of the debate on this topic can be traced back to the uncertainties in the CO-to-H2 conversion 
factor \citep[e.g., ][]{kennicutt12, liu15}. \citet{delosreyes19} argue that this is still a viable prescription for GCE models, so it is reasonable to use the single power-law. Nonetheless, there is sufficient reason to expect the nuances of the star formation law may turn out to be important for chemical evolution. 
}

 \paragraph{The SFR prescription adopted in \GalCEM} 
 We follow the SFR prescription as in \citet{portinari98}:
\begin{equation}\label{eq:SFRdetailed}
\psi(t) = \nu \left[\frac{\sigma(t_G)}{\sigma(t)}\right]^{\kappa-1}G^{\kappa}(t),
\end{equation}
where $G(t) = \sigma_g(t)/\sigma(t_G)$ is the normalized surface gas density:

\begin{align} \label{eq:SFRconversion}
\begin{split}
\psi(t) &= \nu \left[\frac{\sigma(t_G)}{\sigma(t)}\right]^{\kappa-1}\left(\frac{\sigma_g(t)}{\sigma(t_G)}\right)^{\kappa} = \nu \frac{\left[\sigma_g(t)\right]^{\kappa}}{\left[\sigma(t)\right]^{\kappa-1}\sigma(t_G)} \\
&= \nu\frac{\left[M_{\rm gas}(t)/V(t)\right]^{\kappa}}{\left[M_{\rm tot}(t)/V(t)\right]^{\kappa-1}\left(M_{\rm tot}(t_G)/V(t_G)\right)},
\end{split}
\end{align}
where $V(t)$ is the surface area. $M_{\rm tot}(t_G)=M_{\rm final}$ is the final baryonic galaxy mass. Given that the effective radius of the galaxy does not change with time, $V(t_G) = V(0) = V(t) = \textrm{const}$. Let us define the gas fraction $f_g = \sigma_g(t)/\sigma(t)$, so that: 

\begin{figure}[htpb]
\centering
\includegraphics[width = \columnwidth]{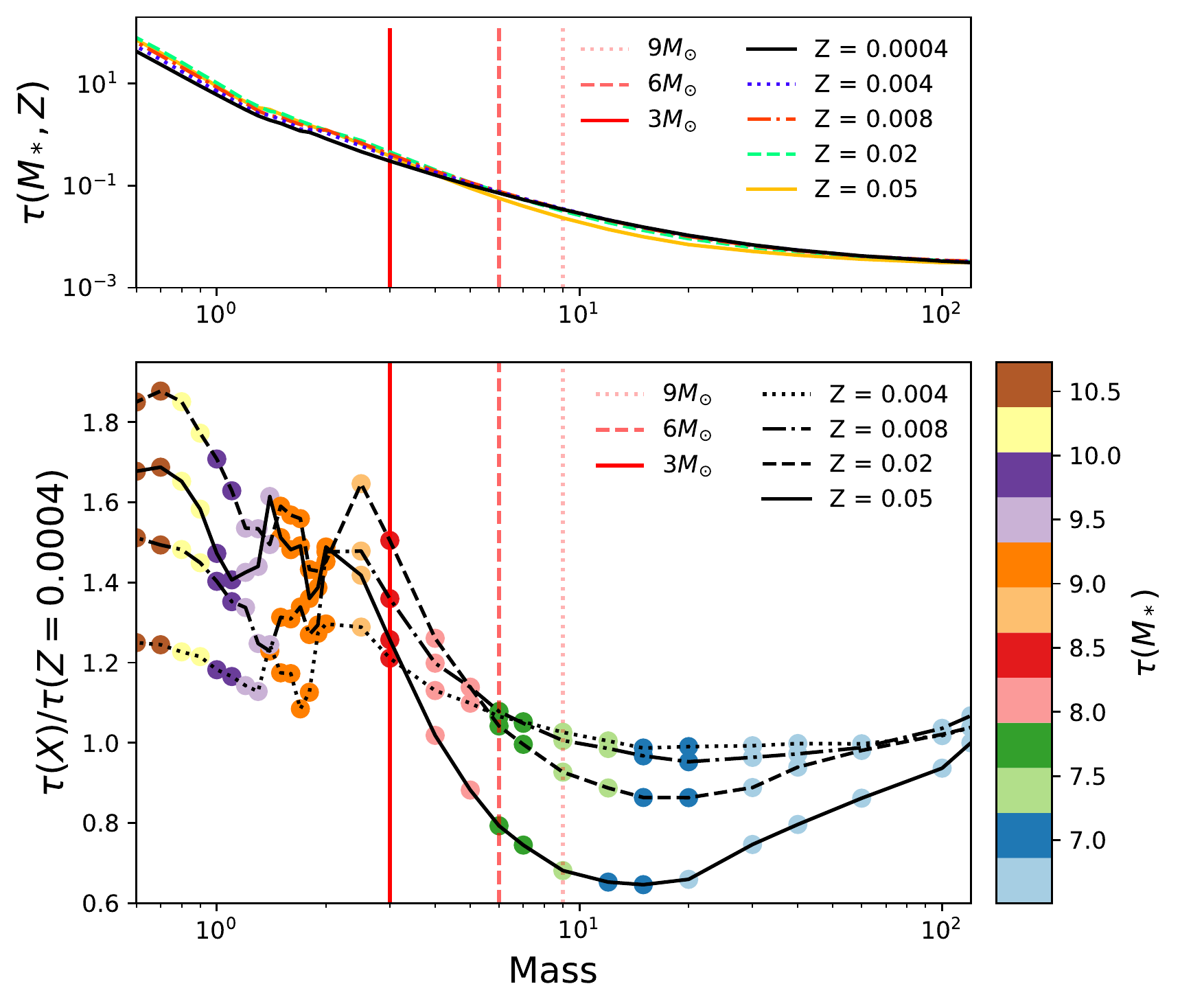}
\caption{The adopted \citep[][]{portinari98} metallicity-dependent stellar lifetimes, $\tau(M_*, Z_*)$, as a function of stellar mass (\emph{top panel}) and the ratio of the top 4 metallicity bins normalized by the lowest one ($Z=0.0004$, \emph{bottom panel}). The 4 next metallicity bins on the bottom panel are $0.004$, $0.008$, $0.02$, and $0.05$ (represented in dotted, dashed, and solid blue lines, respectively). The data points of the metallicity tabulation are color-coded by the actual stellar lifetime as shown in the color bar (where the units are in log$_{10}$ yr). In both the top and bottom panels, three arbitrary stellar masses are highlighted with red vertical lines (3, 6, and 9 \Msun, solid, dashed, and dotted, respectively). The top-left panels in \myfig \ref{fig:stellarlifetimes} represents this same color-coded grid points in a full 3D graphic. } \label{fig:tauratio}
\end{figure}

\begin{align}\label{eq:SFRunits}
\begin{split}
\psi(t) = &\frac{\nu}{M_{\rm final}}\frac{M^{\kappa}_{\rm gas}(t)}{M^{\kappa-1}_{\rm tot}(t)} \equiv \frac{\nu}{M_{\rm final}}\left(\frac{M_{\rm gas}(t)}{M_{\rm tot}(t)}\right)^{\kappa-1}M_{\rm gas}(t) \\
&=\nu f_g^{\kappa-1}(t) \frac{M_{\rm gas}(t)}{M_{\rm final}},
\end{split}
\end{align}
this latter expression being the equation implemented in the code.
Notice that for $\kappa=1$, the SFR is simply proportional to $M_{\rm gas}(t)$. {
The results in this article are presented assuming that $\kappa=1$.}

\subsection{Stellar lifetimes}\label{sec:lifetimes}

Stellar lifetimes is the crucial component that brings together SFR and IMF. Specifically, lifetimes provide the means of connecting stellar masses with time, and they make the integration component of GCE equations possible. 

\paragraph{Stellar lifetimes in \GalCEM}
In the present version of the code, we adopt the metallicity-dependent lifetimes $\tau(M_*,Z)$ from \citet{portinari98}. 
The lifetimes are reported on the top panel of \myfig \ref{fig:tauratio}, where the stellar mass is on the x-axis, the lifetime on the y-axis, and the various curves display varying metallicities. The three red vertical lines aid with tracking 3 reference masses (9 to 3 \Msun)  with lifetimes ranging from $\sim 3\times 10^7$ and $3 \times 10^8$ yr.

The lifetimes in \citet{portinari98} account for the H- and He-burning timescales and are computed with stellar evolution models in the Padua library \citep{bressan93, fagotto94a, fagotto94b}. The computation of these lifetimes also follow the instantaneous mixing approximation, a common assumption in one-zone GCE models. Other burning stages would anyway be shorter than the resolution of these lifetimes \citep[e.g., ][]{limongi17}
\begin{figure*}[htpb]
\centering
\includegraphics[width=\textwidth]{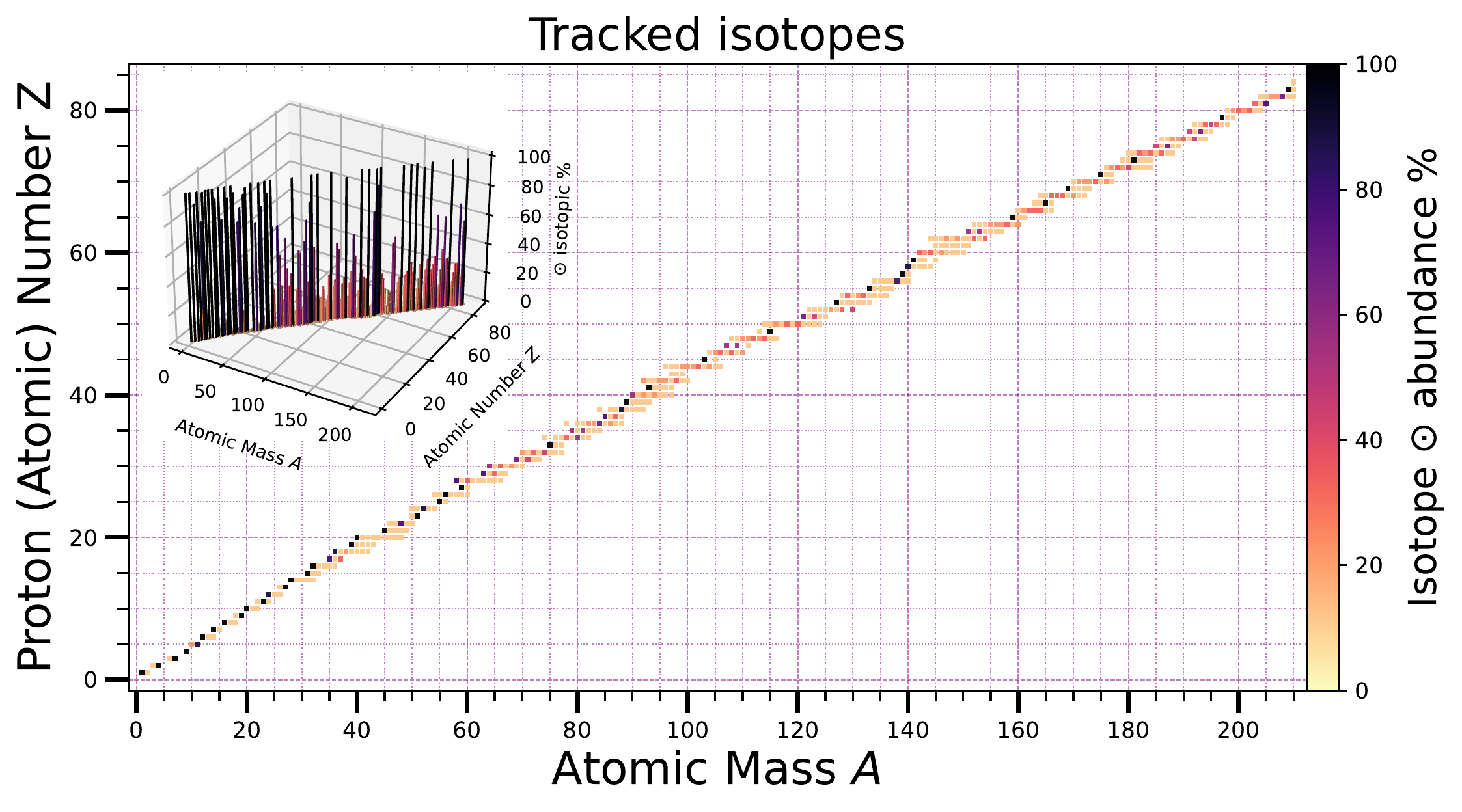}
\caption{All the isotopes (451) tracked in the present run, including the ones coming from LIMs \citep[][]{cristallo15}, massive stars dying as core-collapse SNe \citep[][]{limongi18}, Type Ia SNe \citep[][]{iwamoto99}, and BBN \citep[][]{galli13}. The color bar represents the percentage of each isotope that composes each element in terms of solar abundances \citep[][]{asplund09}. The subplot is a 3D projection of the figure's histogram.}
\label{fig:isotopes}
\end{figure*}
In the bottom panel of \myfig \ref{fig:tauratio} we show the linear ratio between the three lifetimes computed on larger metallicities divided by the lowest metallicity ($Z=0.0004$) lifetime, to better highlight trends. For masses $\gtrsim 6$ \Msun\,and up to solar metallicities,  so no lifetime varies by more than 10\%.

The lifetime at a super-solar metallicity of $Z=0.05$ does not follow the same trends as the ones at lower metallicities. This is caused by the assumed increased relative ratio of He abundance. While in the other lifetimes the dependence on metallicity strongly depends on a higher opacity, for super-solar metallicities two states are at play: a lower hydrogen abundance and a higher average molecular weight $\mu$ -- the latter of which causes a higher luminosity of $L \propto \mu^{7.4}$ \citep{portinari98}. The combination of these two conditions causes the lifetimes to fall for super-solar metallicities in the massive star regime. In the lower mass regime ($\sim 6$ \Msun) the lowest metallicity leads to the shortest lifetime. Having noted these trends, we notice that at all stellar masses, the lifetimes never varies by more than a factor of two. We will show in future works that GCE models are not particularly sensitive to these variations, so that two-Z-bins approaches like \citet[][]{schaller92} or analytical approaches such as \citet[][]{padovani93} are still viable.

\subsection{Yields adopted in \GalCEM}\label{sec:yields}

In \myfig \ref{fig:isotopes} we plot the isotopic solar abundance in the Sun of all the isotopes included in the present run of the simulation, which are 451. We will however limit the discussion to the first 118 isotopes, i.e. from hydrogen to $^{30}$Zn. We will leave the remaining isotopes for the second paper (Gjergo et al., in prep.) where GCE will be explored thoroughly through the introduction of a multitude of r-process enrichment channels.

For LIMs, we submitted a nucleosynthesis query to FUll-Network Repository of Updated Isotopic Tables \& Yields \citep[F.R.U.I.T.Y, ][]{cristallo11,cristallo15}\footnote{Query submitted to \url{http://fruity.oa-teramo.inaf.it/} in July 2021.}. We requested the total isotopic yields for the full stellar mass range ($1.3 \leq M_* \leq 6.0$ \Msun) the full metallicity range ($0.00002 \leq Z \leq 0.02$), a standard $^{13}$C pocket, and all initial rotational velocities, even though in this work we only consider zero initial rotational velocities ($IRV=0$). The F.R.U.I.T.Y. yields were calibrated on \citet{lodders03} solar metallicities, who reported a present-day photospheric metal abundance of $Z=0.0133$. In our paper we adopt \citet{asplund09} with $Z=0.0134$. For consistency, we rescale F.R.U.I.T.Y. to the same metallicities. 

The lowest 4 metallicity bins in F.R.U.I.T.Y. are reported to be $Z=3 \times 10^{-4}$ to $2\times 10^{-5}$. However, these metallicities are $\alpha$-enhanced due to a prevalence of early SN core enrichment. Therefore the authors enhance the $\alpha$ isotopes ($^{12}$C, $^{16}$O, $^{20}$Ne, $^{24}$Mg, $^{28}$Si, $^{32}$S, $^{36}$Ar and $^{40}$Ca) by a factor of 3 ([$\alpha$/Fe]=0.5) and that leads to a metallicity for these lower metallicity bins enhanced by a factor of 2.4 compared to the solar-scaled nominal values. 

{ In \GalCEM\, we leave the raw input untouched, however we follow the repository's instructions: wherever [$\alpha$/Fe] is explicitly indicated, we associate the yields not to the reported metallicity but to a metallicity rescaled by a factor of 2.4.\footnote{"(e.g. the case "0.0001 [$\alpha$/Fe]=0.5" has a metallicity Z=0.00024)", from the (1) HowTo note in \url{http://fruity.oa-teramo.inaf.it/modelli.pl}}}

The SNCC yields are taken from the R set of \citet[][O.R.F.E.O.]{limongi18}\footnote{\url{http://orfeo.iaps.inaf.it/}, set R, tab\_yieldstot\_iso\_exp.dec}. Set R is the recommended set. All stars with $M_*>25$ \Msun\, 
fully collapse into a black hole, so the ejecta in this range come only from the wind component. The mass range is 13 to 120 \Msun\,divided into 9 bins, while there are 4 initial metallicity bins expressed as [Fe/H]=0 to -3. Set R is computed assuming mixing and fall-back, and the mass cut is chosen  so that 0.07 \Msun\, of $^{56}$Ni is ejected for every supernova event. To set up a common baseline, also in this case we consider the zero rotational velocity case. We select the table containing the total final explosive yields with stable isotopes. The assumption on the unstable nuclei is that they fully decay to their closest stable daughter. { We however note that theoretical models of massive stars with physically motivated explosion criteria do not predict a black hole landscape with a simple mass cutoff \citep{ertl16, sukhbold16} as adopted by \citet{limongi18}. This could impact massive star yield ratios based on the mass dependence of explosive yields for different nuclear species.}

The SNIa yields come from \citet{iwamoto99}\footnote{Table 3, formatted in \url{https://github.com/egjergo/GalCEM/tree/main/galcem/input/yields/snia/i99}}, specifically, the favored WDD2 model where the mass of synthesized $^{56}$Ni is set to 0.69 and the explosion energy  to $1.40 \times 10^{51}$ ergs. This model works within a deflagration to detonation transition framework. They assume a central density of $2.12 \times 10^9$ g cm$^{-3}$,  a slow deflagration speed of $v_{def}/v_s =0.015$ after the thermonuclear runaway, where $v_{def}$ is the speed of the deflagration wave while $v_s$ is the local sound speed.

In \GalCEM\, it is possible to select the flags for the selection of the yields or the desired enrichment channel. A constant primordial infall (BBN) is included at setup by default.
 
 \begin{lstlisting}[language=Python]
inputs.include_channel = 
               ['SNCC', 'LIMs', 'SNIa']
inputs.yields_LIMs_option = 'c15'
inputs.yields_SNCC_option = 'lc18'
inputs.yields_SNIa_option = 'i99'
inputs.yields_BBN_option = 'gp13'
 \end{lstlisting}
 
 { The \citet{karakas10} yields have already been processed in \GalCEM\, as well as all the tables from F.R.U.I.T.Y. \citep{cristallo11, cristallo15} and \citet{limongi18}. In the near future we plan to explore the yields computed for stars whose initial rotational velocity is different from zero, and to further expand the yield library to other popular tabulations.}

\subsection{Rates}

Rates return at any given time the newly synthesized isotopes by an enrichment channel in the whole galaxy. In order to carry out the computation, rates reconstruct the occurrence of astrophysical events and pair them with their respective yields. In the case of LIMs and SNCC, the occurrence is marked by the death of individual main-sequence stars. SNIa, on the other hand, are linked to the evolution of a binary system that contains at least one white dwarf. All rates are expressed in units of [\Msun/Gyr].

\subsubsection{LIMs and SNCC rates}

The rates of LIMs and SNCC are given by the following convolution \citep[a historical definition whose first numerical solution dates back to][]{talbot71}: 
\begin{equation} \label{eq:massrates}
    R_{P,i}(t) = \alpha_P \int^{M_{\rm P,u}}_{M_{\rm P,l}} \psi\left(t - \tau_{M_*,Z_*}\right) Y_{P,i, M_*, Z_*} \phi(M_*) \deriv M_*,
\end{equation}
where $\psi$ and $\phi$ are the SFR and IMF respectively. $\alpha_P$ is the fraction of the stars in the integration limits that undergoes the astrophysical process $P$.
For the sake of decluttering, hereafter a subscript indicates a variable dependence (e.g., $\tau_{M_*,Z_*} = \tau(M_*,Z_*)$).

$Y_{P,i,M_*, Z_*}$ are the mass- and metallicity-dependent yields for the $i$-th isotope. The $P$ subscript stands for either LIMs or SNCC. Alternative notations as well as reviews can be found in a variety of sources, including \citet[][]{tinsley80, prantzos08, pagel09, matteucci06}.

A classic convolution has the form $\int^a_b f(t -x) g(x)\deriv x$, while the convolved SFR is a function of time, lifetime, metallicity, and stellar mass, i.e. $\int^a_b f(t -\tau(x,y)) g(x)\deriv x$.
The stellar lifetimes, $\tau_{M_*, Z_*}$, are treated as discussed in Section \ref{sec:lifetimes}. $M_{\rm P,u}$ and $M_{\rm P,l}$ are the upper and lower mass integration limits. The integration is carried with respect to stellar mass. LIMs and SNCC are associated to individual stars. Technically, LIMs integrations should not extend above 6 \Msun\,and SNCC should not fall below 13 \Msun, as those are the largest and smallest bin, respectively, for the F.R.U.I.T.Y. and O.R.F.E.O. yields. 
{Instead of adopting yield computations specific to this mass range, we extrapolate the yields linearly to a mass of 10 \Msun\,on both ends}.

\paragraph{Rates for single-star enrichment channels in \GalCEM}
The integration limits assume that enough time has passed so that stars may die. In \GalCEM\, we impose this condition by ensuring that the birth-time is always positive. Birth-time $t'$ is defined as the difference between galaxy time and stellar lifetime. This is reflected in the integration limits of \myeq \ref{eq:birth timerate} -- after a change of variables that switches the independent variable of the integral $M_*$ to its birth-time $t'$. Birth-time is univocal only to a given stellar population.

In the case of LIMs, there will be an overlap with the binary white-dwarf progenitor that will produce SNIa (the very next Section). By defining $\alpha_{SNIa}$ the fraction of binaries that will produce SNIa, the LIMs rate which overlap the SNIa rate will be rescaled by a factor $\alpha_{LIMs}= 1 - \alpha_{SNIa}$.

\subsubsection{Type Ia supernova rates}
There is an extensive body of literature investigating SNIa rates \citep[for a review, see][]{maoz12}. SNIa have been shown to be responsible for the production of about $2/3$ of the Fe content of a galaxy \citep{matteucci05}, although we note that this fraction is sensitive to the adopted IMF. { SNIa may occur either when a white dwarf accretes mass from a close binary companion (single-degenerate scenario, SD) or when two white dwarfs in a binary system coalesce (double-degenerate scenario, DD). \citet{GonzalezHernandez12} finds that the SD scenario should not occur in more than 20\% of the SNIa events.  Studies on solar neighborhood abundances do not show a strong preference for either a SD or DD scenario \citep{matteucci09}. What they do require is a large delay, i.e., of the total number of SNIa, a fraction not larger that 30\% (and preferably smaller than 20\%) should have occurred within timescales shorter than 100 Myr \citep[][]{matteucci06, matteucci09}. Even larger delay times are predicted in \citet[][]{totani08} but in that prescription, a double-degenerate scenario is favored. The necessity for significant delay times is also apparent through other empirical findings. For example, \citet{holoien19} finds that in the ASAS-SN bright supernova catalog, over 10\% of the events occurs over 10 kpc away from their host galaxy, suggesting that the binary progenitors migrated considerable distances before exploding.}

\paragraph{SNIa rates in \GalCEM}
{ For SNIa rates we follow the SD scenario proposed by \citet[eq. 16,][]{greggio05}, which is an improvement on the SD models employed in \citet{greggio83} and \citet{matteucci86}. 

In this scenario, the SNIa rate is given by: 
\begin{multline}\label{eq:Rate_SNIa}
    R_{\rm SNIa,i}(t) = \alpha_{\rm SNIa}\; Y_{\rm SNIa,i}\\
    \int^{{\rm{min}}(t,\tau_{\rm x})}_{\tau_{\rm i}} \psi(t-\tau)DTD_{\rm SNIa}(\tau) \deriv \tau,
\end{multline}
where $\alpha_{SNIa}$ absorbs $k_{\alpha}$, the number of stars per unit mass in one stellar generation, which is equivalent to 1.55 for the \citet{kroupa01} IMF, as well as the realization probability of the SNIa scenario $A_{SNIa}$, set to $10^{-3}$, according to \citet{greggio05}, for our adopted canonical IMF. The yields $Y_{\rm SNIa}$ are assumed to not vary as a function of time or mass. The delay time distribution of SNIa, $DTD_{\rm SNIa}$, after being normalized to 1 ($\int^{\tau_{\rm x}}_{\tau_{\rm i}}DTD_{\rm SNIa}(\tau){\rm d}\tau=1$) is then convolved with the SFR $\psi$ and integrated across delay times. $\tau_{\rm i}$ corresponds to the lifetime of a progenitor of $\sim 8 M_{\odot}$, i.e., the most massive star capable of producing a WD. In the case of the SD model, $\tau_{\rm x}={\rm min}(m_{2,e})$ is set by a limit on, the envelope mass $m_{2,e}$ of the mass of the secondary component of the binary system, $m_2$\footnote{$m_2$ is the companion with a smaller mass -- $m_2 \leq m_1$ where $m_1$ is the primary, more massive companion.}, which should not fall below $m_{2,e}> 0.15/\epsilon$. $\epsilon$ is an efficiency parameter that determines how much of the mass of the secondary companion is accreted onto the WD, and it will appear again at the end of this subsection.

The $DTD_{\rm SNIa}$ in the SD scenario is proportional to two quantities that depend on $m_2$. Specifically, it is proportional to the absolute value of the time derivative of the mass, $|\dot{m}_2|$, and to the distribution function of the secondary in a progenitor system, $n(m_2)$ :
\begin{equation}
    DTD_{\rm SNIa} \propto n(m_2)|\dot{m}_2|.
\end{equation}

The mass of the secondary is well approximated by the main-sequence lifetime $\tau_{\rm MS}$  by \citet{girardi00} in the mass range $0.8 \lesssim m_2/M_{\odot} \lesssim 8$, corresponding to $0.04 \lesssim \tau_{\rm MS}/{\rm Gyr} \lesssim 25$:
\begin{equation} \label{eq:lifetimesecondary}
    \log m_2 = 0.0471 \left(\log \tau_{\rm MS}\right)^2 - 1.2 \log \tau_{\rm MS} + 7.3.
\end{equation}

The distribution function of the secondary mass is given by an integral over the mass of the primary mass, and will depend on the slope of the power law distribution of the binary system, $-\alpha$, and on the slope of the power law distribution of the ratio between secondary and primary masses, $\gamma$. The integral simplifies to \citep[eq. 16][]{greggio05}:
\begin{equation}
    n(m_2) \propto m_2^{-\alpha} \left[ (m2/m_{1,i})^{\alpha+\gamma}) - (m_2/8)^{\alpha+\gamma}\right],
\end{equation}
where $m_{1,i}$ is the minimum mass of the primary companion. It is constrained ($m_{1,i}={\rm max}(m_2, m_{1,n})$) so that it is the largest between $m_2$ and the remnant
mass of the primary ($m_{1,n}= {\rm max}\left\{2., 2.+10.(m_{\rm WD,n}-0.6)\right\}$), which is constrained by the minimum acceptable mass for the WD, i.e.
$m_{\rm WD,n} =1.4 + \epsilon \, m_{2,e}$. In this work we take $\epsilon=1$, i.e., the solid curves in Fig.2 of \citet{greggio05}. 

The mass of the envelope of the secondary, $m_{2,e}=m_2-m_{2,c}$, is constrained by its own secondary remnant mass, derived in \citet{nelemans01} to be: 
\begin{equation}
    m_{2,c} = {\rm max}\left\{0.3; 0.3+0.1(m_2-2); 0.5+0.15(m_2-4) \right\}.
\end{equation}
}

\begin{figure*}[htpb]
    \centering
    \includegraphics[width = \textwidth]{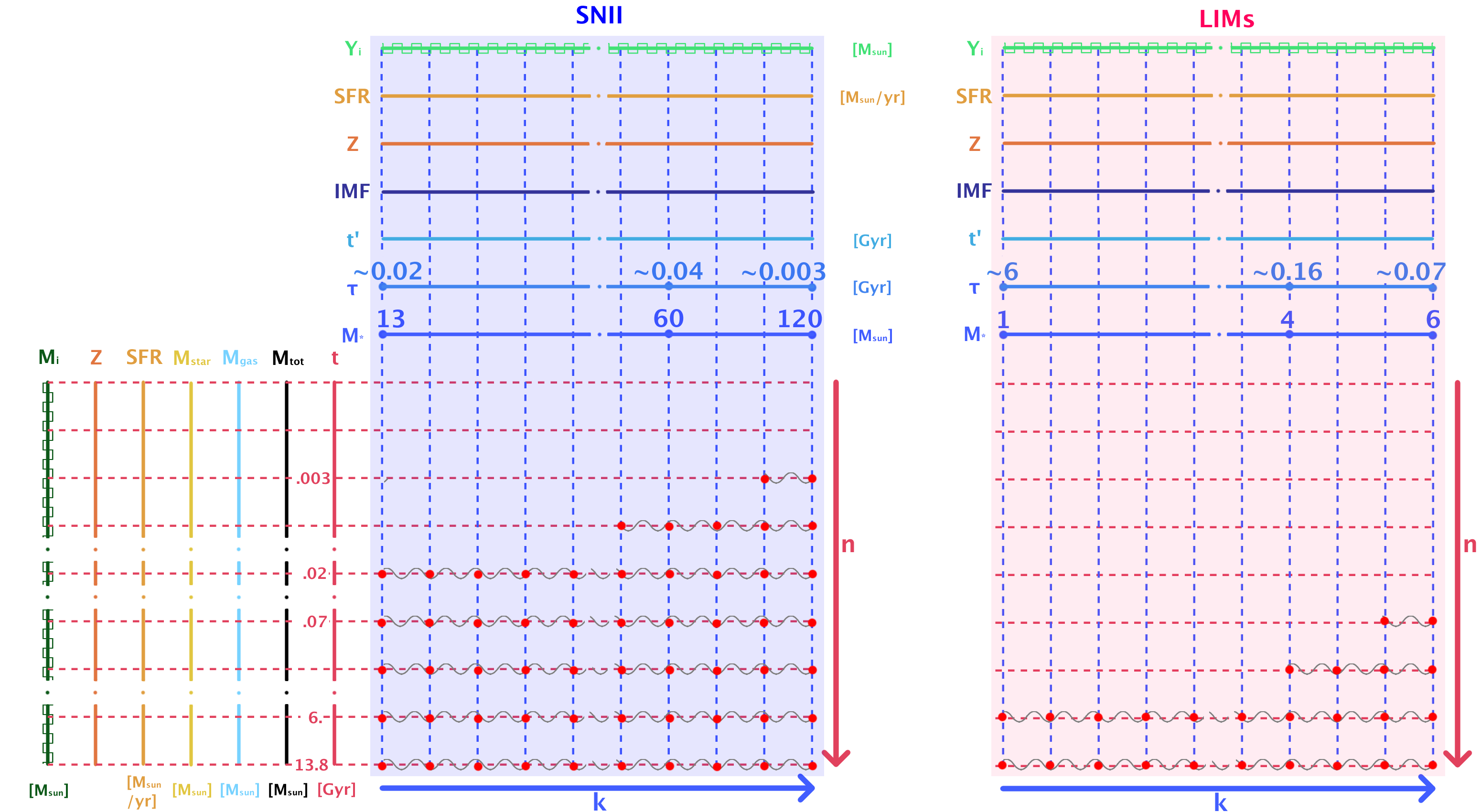}
    \caption{Schematic diagram representing the solution to the integrodifferential GCE equation. A description of the diagram can be found in Section \ref{sec:integr}}
    \label{fig:integrGrid}
\end{figure*}

\subsection{General integrodifferential GCE Equation}\label{sec:general}

The full GCE equation solved in \GalCEM\, is given by \myeq \ref{eq:general}. And it expresses the rate of change of the gas mass of isotope $i$ in units of [\Msun/Gyr].

\begin{multline}\label{eq:general}
\dot{M}_{\rm i,gas}(t) =  X_{\rm i,inf}\dot{M}_{\rm inf}(t)  -(1-\omega)X_{\rm i,gas}(t) \psi(t) \\
  + \sum_P \alpha_P R_{P,i}(t)
\end{multline}

The first term on the RHS is the infall component, while the second term is the SFR and outflow 
component, with $\omega$ being the wind efficiency of the outflow. The third and last term is the summation of the rate integrals of all the enrichment channels for the isotope $i$, with the rate expressed in full in \myeq \ref{eq:birth timerate}. $\alpha_P$ is the fraction of the stars in the enrichment channel mass range which will undergo the given astrophysical event.

The first and second terms (RHS on the first row of \myeq \ref{eq:general}) are rescaled to the respective isotopic gas mass fractions $i$ ($X_{\rm i,inf} = M_{\rm i,inf}/\sum_i M_{\rm i,inf}$), which in the first term reflects the primordial composition of the infall gas. In the second case, by applying the instantaneous mixing approximation, the fraction $X$ refers to the $i$th gas mass fraction as a function of time $t$, $X_{\rm i,gas}(t) = M_{\rm i,gas}(t)/\sum_i M_{\rm i,gas}(t)$. 

In the third term, the summation of the enrichment channel rates, a change of variables has been applied, so that the integral is expressed in terms of birth-time, $t' = t - \tau_{M_*,Z_*}$.
The following rate (or a variation thereof) applies to any enrichment channel in which the yield is either mass or metallicity dependent:
\begin{align}\label{eq:birth timerate}
\begin{split}
R_{P,i}(t)=&
\int^{t-\mathrm{min}(t,\tau_{(M_u,Z_{M_u})})}_{t-\mathrm{max}(t,\tau_{(M_l,Z_{M_l})})} 
 \textrm{d}t' \psi(t') \times \\
 & \bigg\{{-\frac{\textrm{d}M_*\left(t-t',{Z_{t'}}\right)}{\textrm{d}\tau}} 
 {\phi\left[M_*(t-t',{Z_{t'}})\right]} \\ &{Y_{P,i}\left[M_*\left(t-t',{Z_{t'}}\right)\right]}  \bigg\}_{M_*\left(t-t',{Z_{t'}}\right)},
\end{split}
\end{align}

 \begin{figure}[htpb]
\centering \includegraphics[width = \columnwidth]{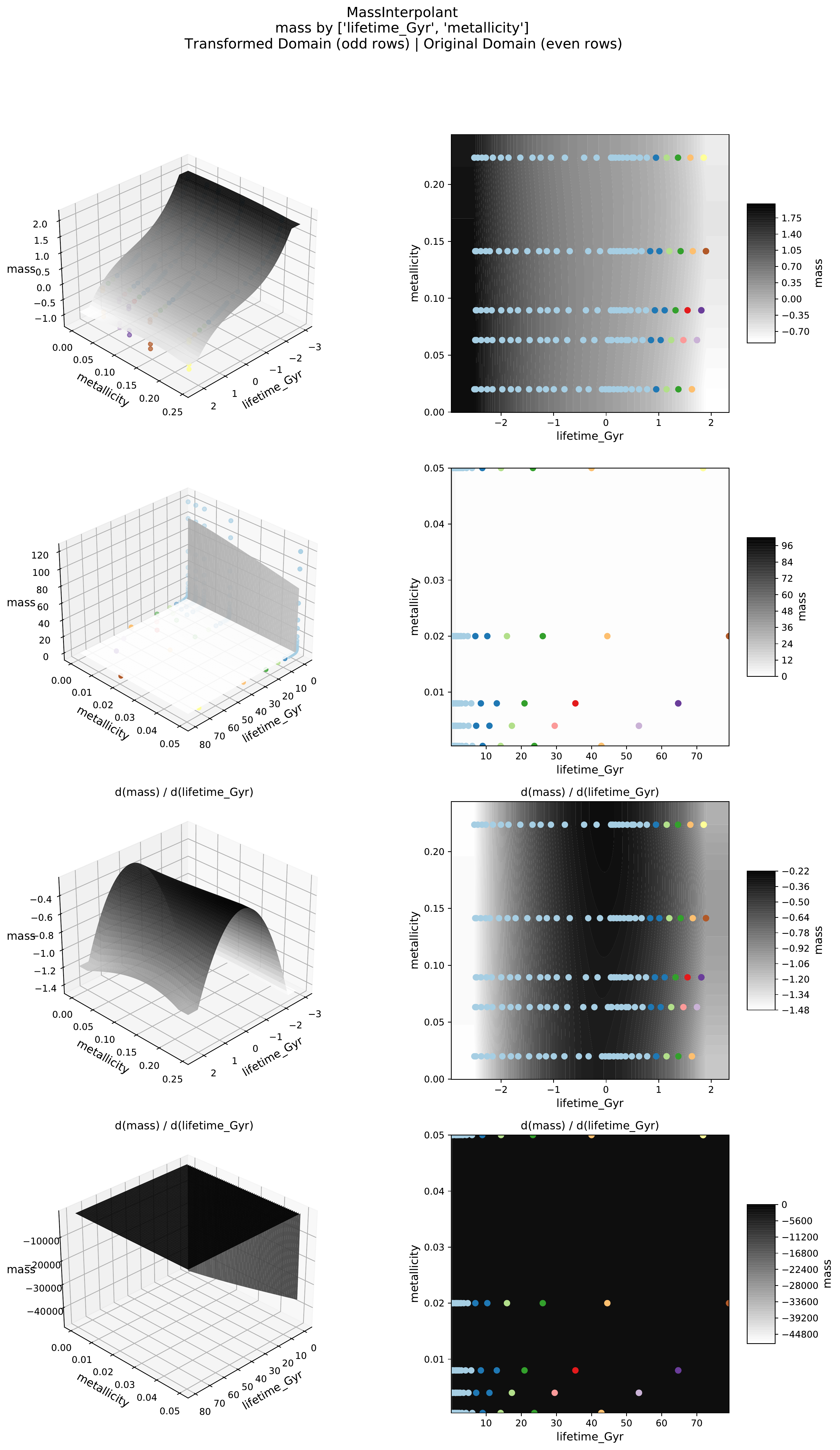}
\caption{Mass interpolant employed in this work and explained in 
Section \ref{sec:lifetimes}. The even rows represent the original domain, and the odd rows the transformed domain onto which the interpolation is computed. The top half of the image shows the actual mass interpolant, while the bottom half shows the derivative of the mass with respect to the lifetime. The 3D projections in the first column show the metallicity and lifetime in Gyr for the bottom x-y plane, while mass (or lifetime derivative of the mass for the bottom panels) is shown on the vertical z-axis.  The second column displays the respective 2D contours. The color-coding of the scatter points is consistent with \myfig \ref{fig:tauratio}.
}\label{fig:stellarmasses}
\end{figure}

 \begin{figure}[htpb]
\centering \includegraphics[width = \columnwidth]{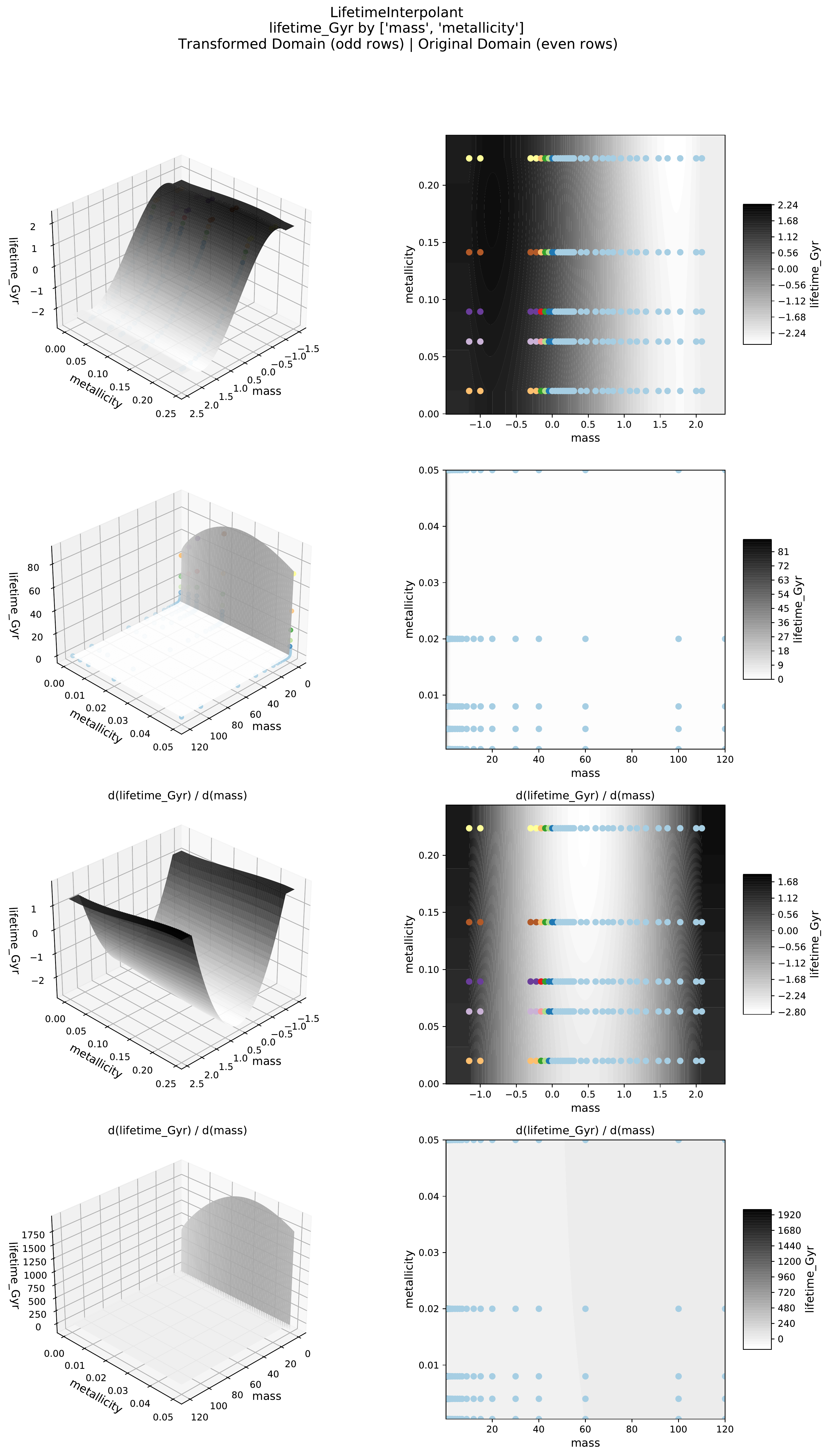}
\caption{Lifetime interpolant employed in this work and explained in Section \ref{sec:lifetimes}. It follows the same structure as \myfig \ref{fig:stellarmasses}, but for $\tau(M_*, Z)$; i.e., the even rows represent the original domain, and the odd rows the transformed domain onto which the interpolation is computed. The top half of the image shows the lifetime interpolant, while the bottom half shows the derivative of the lifetime with respect to the mass. The 3D projections in the first column show the metallicity and lifetime in Gyr for the bottom x-y plane, while mass is shown on the vertical z-axis. In the second column are the respective 2D contours. The color-coding of the scatter points is consistent with \myfig \ref{fig:tauratio}. }\label{fig:stellarlifetimes}
\end{figure}

 \begin{figure*}[htpb]
 \includegraphics[width = \textwidth]{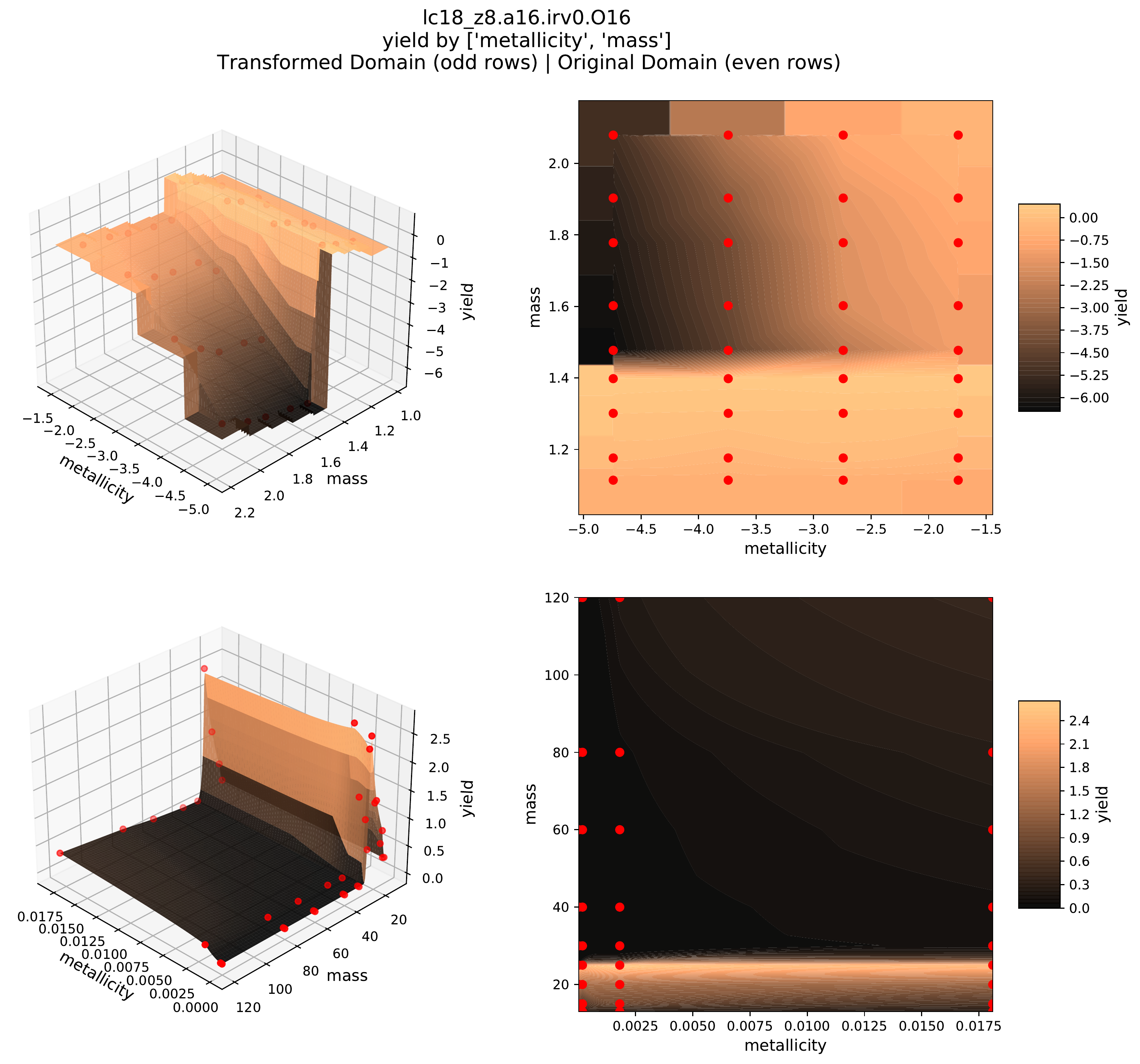}
\caption{Example of interpolating yield tabulations. The current plot refers to $^{16}$O for the massive yields by \citet{limongi18}. The two plots on the bottom show the fit to the raw data. On the top is shown the interpolation on the transformed domain. This is the preprocessed interpolation curve which is read inside the SNCC rate integral. An equivalent computation for LIMs is shown in \myfig \ref{fig:o16interp15}.} \label{fig:o16interp18}
\end{figure*}

 \begin{figure*}[htpb]
\centering \includegraphics[width = \textwidth]{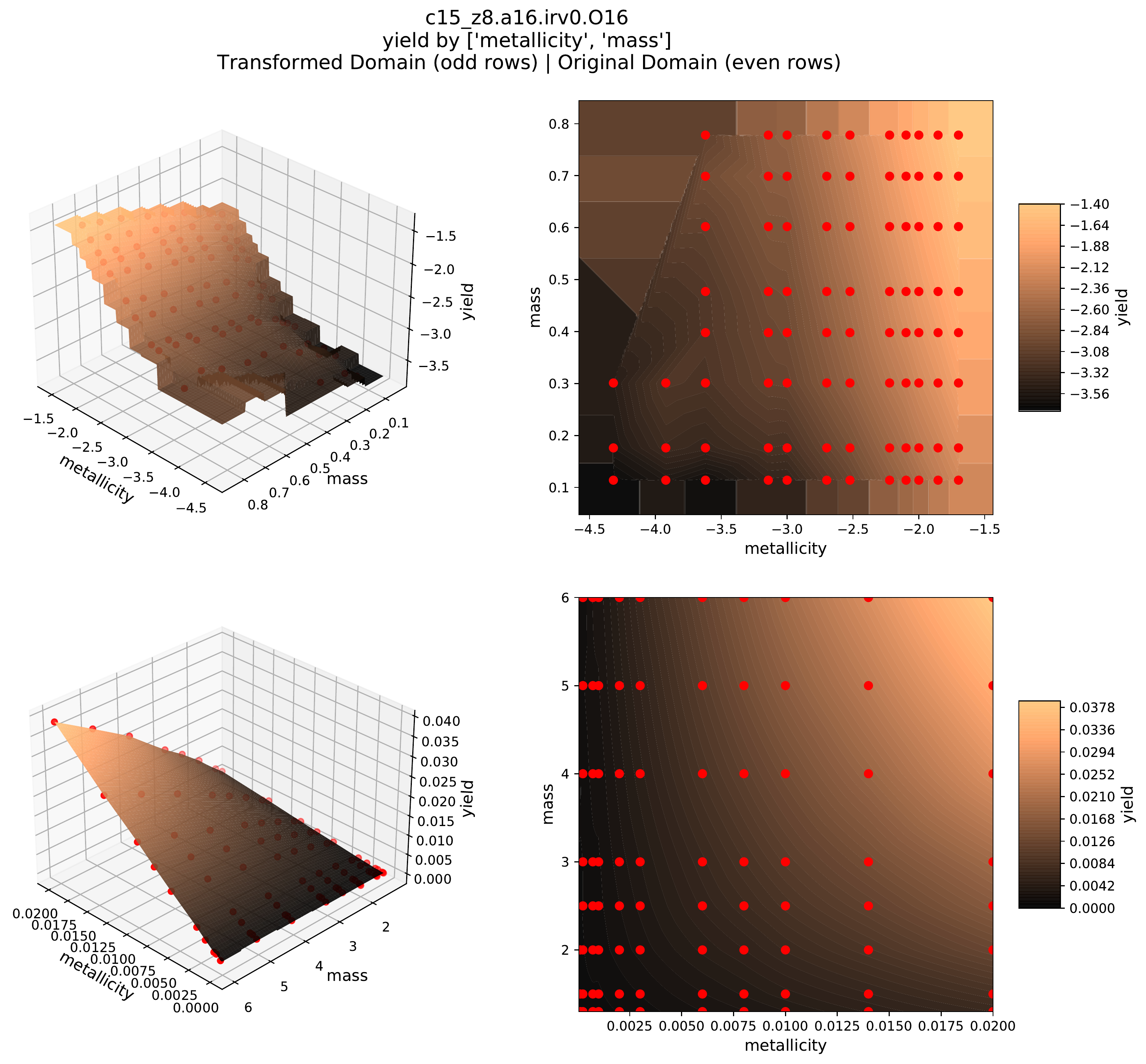}
\caption{Example of interpolating yield tabulations. The current plot refers to $^{16}$O for the LIMs yields by \citet{cristallo15}. Similarly to \myfig \ref{fig:o16interp18}, the bottom plot shows the interpolation on the raw data while the top plot shows the interpolation on the transformed domain. This latter curve is the preprocessed interpolation which is read inside the LIMs rate integral.}\label{fig:o16interp15}
\end{figure*}

\noindent where the integral is computed in the  range [$t'(M_{\rm l,P}),t'(M_{\rm u,P})$], which undergoes the astrophysical process $P$.  $t'(M_{\rm l,P})= t-\mathrm{max}(t,\tau_{(M_l,Z_{M_l})})$, and $t'(M_{\rm u,P})=  t-\mathrm{min}(t,\tau_{(M_u,Z_{M_u})})$. That is to say, the mass limits of integration of \myeq \ref{eq:massrates} correspond to the respective birth-time limits $t'(M_{\rm l,P})$ and $t'(M_{\rm u,P})$. This prescription is consistent with the \citet{matteucci86} formalism, later expressed explicitly in \citet{portinari98}.

The integral will be computed when (1) more time has passed than the lifetime of the most massive star, and (2) the birth-time is positive, i.e. there are stars which have died at time $t$.
The integrand terms within the curly brackets of \myeq \ref{eq:birth timerate} are the stellar-mass-dependent convolution pulses. The terms consist of the IMF, $\phi(M_*, t-t', Z_{t'})$, the yields $Y_{P,i}$ for any given enrichment channel $P$ and isotope $i$, and a chain rule element emerging from the change of variables from $M_*$ to $t'$\footnote{The chain rule reads:\\$
    \deriv M_*=\frac{\deriv M_*}{\deriv t'} \deriv t' = \frac{\deriv M}{\deriv \tau} \frac{\deriv \tau}{\deriv t'} \deriv t' = - \frac{\deriv M_*(\tau_*,Z_*)}{\deriv \tau(M_*,Z_*)}\deriv t',$}.
The negative sign emerges due to $\deriv \tau/\deriv t'=-1$.
The integral is computed as a function of stellar birth time $t'$, where $t' = t - \tau(M_*,Z_*)$. Elsewhere in this article written as $\tau_{M_*,Z_*}$, the lifetime is the function described in Section \ref{sec:lifetimes} that returns the span of existence in Gyr of a star of mass $M_*$ and metallicity $Z_*$.  Inside the integral, the metallicity of the star $Z_*$ is reconstructed from the Galaxy metallicity $Z(t)$ via the lifetime of the star of mass $M_*$. So the term $\deriv M_*$ can be converted to $\deriv t'$ via the first order derivative of the stellar mass with respect to the stellar lifetime.

\subsubsection{Integration of the Rates in \GalCEM} \label{sec:integr}
The whole stellar mass range has to be split appropriately to isolate only the stars associated with the yield $Y_P$. The mass range can be denoted by the stellar limits $M_{\rm l,P}$ and $M_{\rm u,P}$ for the lower and upper mass limit respectively.
In \GalCEM\, all of the components of the integrand are evaluated on a mass grid constructed between $t'(M_{\rm l,P})$ and $t'(M_{\rm u,P})$ on a grid of size $N_k$ with $k=200$ \citep{portinari98}, uniformly distributed in log-space. 

At each time step and for every process, \GalCEM\, computes a grid of quantities starting from the stellar mass, the lifetime, and the birth-time. Onto these grid points it evaluates all the functions in the integrand. This includes for example the SFR, $\psi(t')$, which is evaluated with respect to the birth-time: i.e., the past $\psi(t)$ is interpolated in order to reconstruct the SFR at the time of birth of all the stars that die at any given time step.

\begin{figure*}[htpb]
\centering
\includegraphics[width = \textwidth]{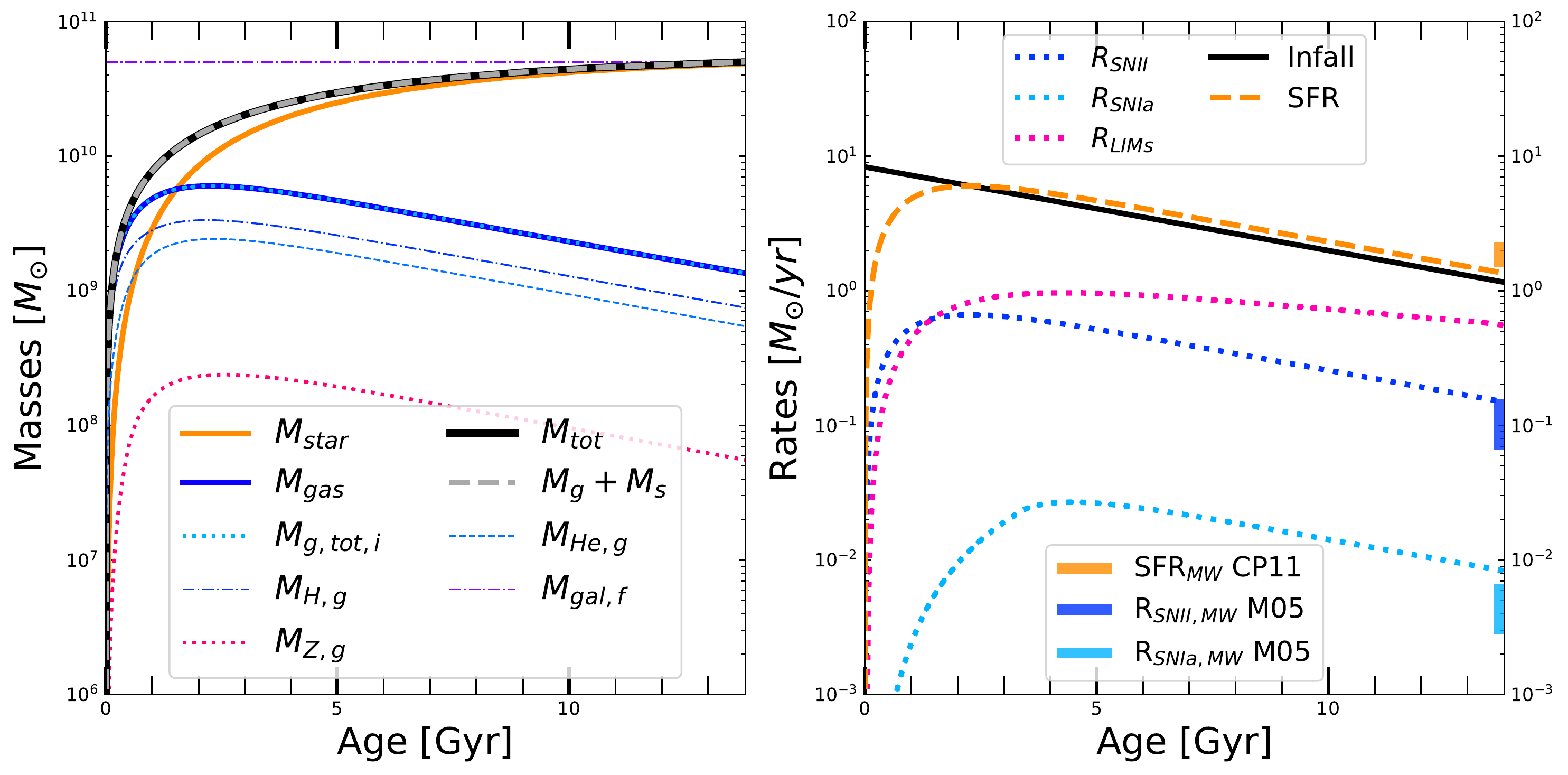}
\caption{Evolution of the global quantities in a GCE solution. The figure on the left represents mass quantities, while the figure on the right represents rate quantities. $M_{\rm tot}$ is the total time-dependent baryonic mass. $M_{\rm star}$ is the total stellar mass, or $M_{\rm *,tot}$ in the text. $M_{\rm gas}$ is the total gas mass. The sum of total stellar and total gas mass, $M_{\rm gas}+M_{\rm *,tot}$ as a sanity check coincides with $M_{\rm tot}$. $M_{\rm g,tot,i}$ is the total gas mass as inferred from the isotopic evolution output. As another sanity check, it should coincide with $M_{\rm gas}$. $M_{\rm H,g}$, $M_{\rm He,g}$, and $M_{\rm Z,g}$ are the time-dependent masses of H, He, and all metals, respectively. $M_{\rm gal,f}$ represents the final baryonic mass of the galaxy. When it comes to the rates, the infall is shown in the black solid line, the SFR is in the dashed orange yellow line, and the dotted dark blue, light blue, and magenta lines represent the SNCC, SNIa and LIMs rate respectively. The present-day Milky Way estimates of the rates are shown with the vertical segments at time 13.7 Gyr. The SFR is taken from CP11, \citet{chomiuk11}, while the SNCC and SNIa rates come from M05,  \citet{mannucci05}.} \label{fig:totphys}
\end{figure*}

\myfig \ref{fig:integrGrid} is a schematic representation of the  grid across which the returned rate integral is computed. Galaxy age $t$ grows vertically downwards and has a size of $n$, while stellar mass $M_*$ is shown horizontally on a grid of length $k$. At each time step, there are a series of stored total physical quantities (whose growth is plotted in \myfig \ref{fig:totphys}). From right to left there is galaxy age $t$, total baryonic mass $M_{\rm tot}(t)$, total gas mass $M_{\rm gas}(t)$, total stellar mass $M_{\rm star}(t)$, SFR$(t)$, metallicity, and finally in dark green is the ($N_i,N_n$) matrix representing the mass [\Msun] of every isotope as a function of time. The boxed line indicates $M_i(t)$ is a matrix instead of a vector like the other quantities.  On the horizontal axis are the integral components for every enrichment channel, evaluated at every time step and for every isotope. From bottom to top are the stellar mass $M_*$, the stellar lifetime, the stellar birth-time, the IMF, the metallicity evaluated at the stellar birth-times, $Z(t')$, and similarly the SFR, $\psi(t')$, is also evaluated at the stellar birth-times, and finally the yield interpolations $Y_i$. There are two quadrants shaded in blue and pink that represent the SNCC and LIMs enrichment channels, respectively. The only independent variables are $t$ and $M_*$.

For each time step and for each enrichment channel, an integral must be solved that returns the total yield from all the events that occur at that given time step. On the vertical axis are the time-dependent quantities emerging from the solution of the differential equations.  We label a selection of grid points, with a sampling that is not to scale. We have also shown plausible lifetimes for the mass grid, evaluated on the metal poor regime.  An enrichment channel will activate only when the galaxy time elapsed will be longer than the shortest lifetime of the channel's stars, and the $N=200$ mass grid will be cropped to exclude the lower mass stars which have not died yet.

The analytical form of the GCE integrands, $F_i$, have the following form:
\begin{equation}
    F_i(t,M_*) = \psi(t-\tau(M_*)) \times \left[Y_i,\phi\right](M_*)
\end{equation}
with the change of variables we transform them into:
\begin{equation}
    F_i(t,t') = \psi(t') \times \left[-\frac{\deriv M_*}{\deriv \tau}, Y_i, \phi \right]\left(M_*(t-t')\right),
\end{equation}
i.e., the stellar-mass-dependent component is a function dependent on galaxy age, stellar birth-time $w(t,t') = \left(
\deriv M / \deriv \tau \times \phi \times \textrm{Y}_i\right) \left(M_*(t - t') \right)$.

\begin{figure}[htpb]
    \centering
    \includegraphics[width=\columnwidth]{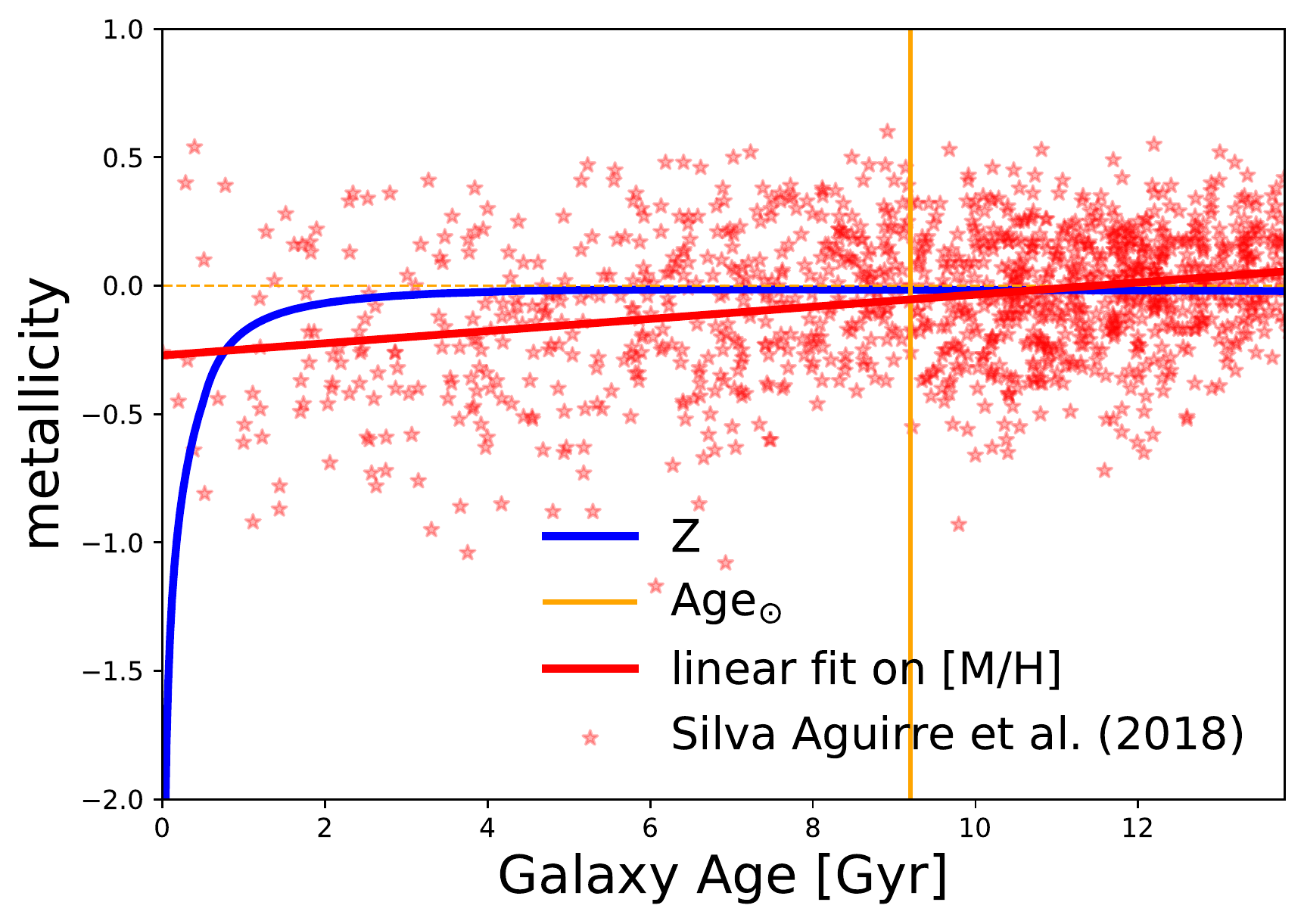}
    \includegraphics[width=\columnwidth]{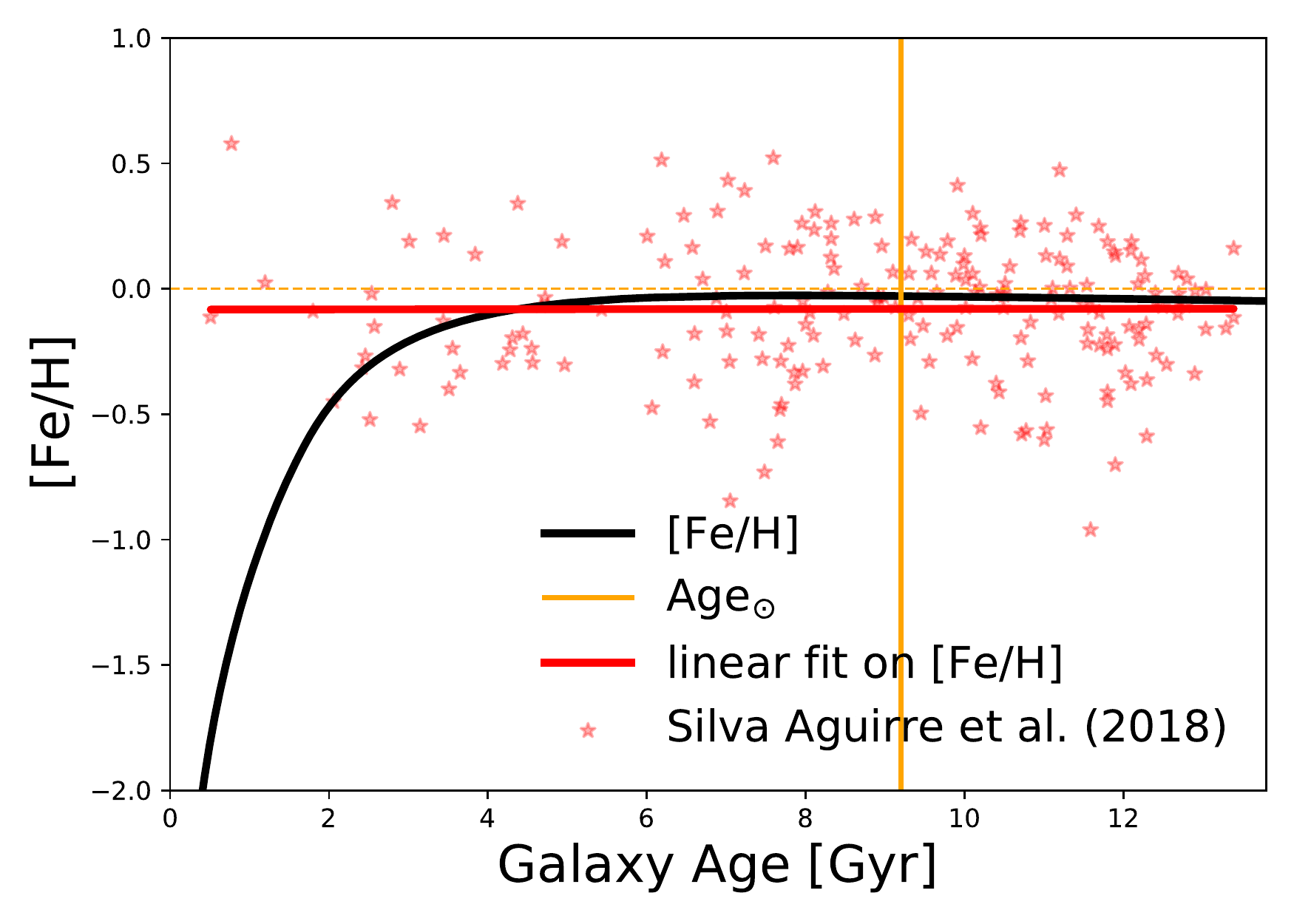}
    \caption{The age-metallicity relation with respect to the metallicity (upper figure) and iron abundance (lower figure) normalized to solar values. The orange lines cross at the solar age and abundance. The observational scatter comes from the APOKASC sample. Stellar age and abundances are taken from \citet{silvaAguirre18}, while the iron abundance is taken from \citet{pinsonneault14}.}
    \label{fig:ageZ}
\end{figure}

Having mapped appropriately each item in this integrand to its appropriate grid point as shown in \myfig \ref{fig:integrGrid}, the integrand is simply given by the following product:
\begin{equation}
F\left(t, t'\right) =\psi\left(t'\right)   w(t, t').
\end{equation}
So any convenient integration method compatible with the Volterra Equations is capable of dealing with the integrand $F\left(t, t'\right)$. We apply the Simpson's rule.

\paragraph{Solving the Differential Equation in \GalCEM} \label{sec:diffeq}
At each time step, the integrals of each isotope and each enrichment channel is summed into a single quantitity to be added to the total gas mass component. To solve the differential equation we simply solve a classic fourth-order Runge-Kutta method to the total galaxy quantities, namely SFR, stellar mass, and gas mass. At this stage also the metallicity is updated.

{Currently, \GalCEM\, runs on a uniform time step. The output of the runs presented in this paper results form a $\Delta t = 2$ Myr. Adaptive time steps will be tested in the future to improve the computation times.}
 
\subsection{Summary of the {\tt GalCEM} framework} \label{sec:numeric}
With the yield selection outlined in Section \ref{sec:yields}, \GalCEM\, runs on 451 isotopes $i$ for 86 chemical elements.
The variable $t$ represents the age of the galaxy.
By solving for $M_{\rm i,gas}(t)$ in \myeq \ref{eq:general}, the goal is to obtain an ($i, t$) matrix as output, with each entry representing the mass of every isotope as a function of time. \myeq \ref{eq:birth timerate} is, for each enrichment channel, a system of equations of size $i$, computed at every time step.
$\dot{M}_{\rm inf}(t)$ is \myeq \ref{eq:infall}, fully and uniquely defined by the parameter $\tau_{inf}$, chosen at the start of the run, therefore we compute both $\dot{M}_{\rm inf}(t)$, the infall rate, and ${M}_{\rm inf}(t)$, the total mass of the system at $t$ at the setup stage.
$\dot{M}_{\rm *,tot}(t)$ is the SFR defined in \myeq \ref{eq:SFR}. It depends on $\dot{M}_{\rm inf}(t)$ (\myeq \ref{eq:infall}) and on $M_{\rm gas}(t)=\sum_i M_{\rm i,gas}(t)$ (that we are solving for). we save both the one-dimentional vector of length $t$ for $\psi(t)=\dot{M}_{\rm *,tot}(t)$, the SFR, and ${M}_{\rm *,tot}(t)$, the total stellar mass of the system at $t$.

There is not a single lifetime relation $\tau_Z(M_*)$, in \citet{portinari98} there are 4, as seen in Fig. \ref{fig:stellarlifetimes}, depending on the initial metallicity of the star ($Z = 0.05, \, 0.02, \, 8 \times 10^{-3}, \, 4 \times 10^{-4}$). These functions are interpolated and extrapolated in accordance with the method described in the following section, Section \ref{sec:interpolationtool}.

The convolved integral is given by the product of two functions: the SFR as a function of birth time $t'$, and a product of quantities that depend on the lifetime and metallicity-dependent stellar mass $M_*(t-t'_{Z_{t'}})$.

\subsection{{\tt GalCEM} interpolation tool} \label{sec:interpolationtool}
Solving detailed GCE equations requires interpolating at several stages: the yield tabulations must be interpolated over mass, metallicity, and occasionally other parameters such as initial rotational velocity. Inside the integrals, also stellar lifetime, metallicity, and SFR need to be interpolated. Lastly, an interpolation is required for the derivative of the stellar mass with respect to stellar lifetime. Yields, mass, and derivatives can be preprocessed with the {\tt GalCemInterpolant} class in the {\tt yield\_interpolation} package. 

With the aim of computing the derivative of the stellar mass with respect to its lifetime, we derive the fits as they appear in \myfig \ref{fig:stellarmasses} where lifetimes and metallicities are interpolated to obtain stellar masses, and \myfig \ref{fig:stellarlifetimes} where masses and lifetimes are interpolated to obtain lifetimes\footnote{The interpolation code can be found in \url{https://github.com/egjergo/GalCEM/tree/main/yield_interpolation/lifetime_mass_metallicity/main.py}}. We implement the {\tt SciPy} \citep[][]{scipy20} BivariateSpline interpolation which, while not providing as good of a fit like other methods such as LinearNDInterpolator and NearestNDInterpolator, is smooth so it supports taking derivatives. Bivariate splines are ideal methods for the solution of boundary-value problems by finite-element-type methods because they consist of piecewise polynomials triangulated on a polygonal domain \citep[e.g.,][]{nurnberger00}.

The even rows of both \myfig \ref{fig:stellarmasses} and \myfig \ref{fig:stellarlifetimes} are shown in the transformed domain. The transformations implemented by the interpolator are: logarithmic mass, square root metallicity, and logarithmic lifetime. The odd rows depicts the original domain. The transformation was necessary to ensure stable fits, especially with the derivatives and specifically the $d \tau/d M_*$ fit which we employ in the returned rate integral. 

\myfig \ref{fig:o16interp18} and \myfig \ref{fig:o16interp15} are the interpolation surfaces computed for SNCC and LIMs, respectively. Both the original and the transformed domain are shown in order to display the fit comparison. The statistics for the goodness of the fit are reported in Section \ref{sec:appendix}. 
Unlike \myfig \ref{fig:stellarmasses} and \ref{fig:stellarlifetimes}, \myfig  \ref{fig:o16interp18} and \ref{fig:o16interp15} are obtained with a combination of LinearNDInterpolator, NearestNDInterpolator. These two methods together give better recovery of known values. However, the models are less smooth, especially outside the hull where the derivative is 0 almost everywhere and undefined at points equidistant between neighbors. 
{Bivariate splines do not recover y values at fitted x values while the linear nd interpolator, the nearest neighbor does. This comes at the cost of lacking smoothness. The linear nd interpolator makes predictions for points inside the hull and the nearest neighbor outside
. The errors in Listings \ref{c15output} and \ref{lc18output} in the Appendix are computed with the data used to fit models (in-sample)
. The original domain plots  shows predictions from the model fit in the transformed domain i.e. the models are not separate, rather they are viewed in different scalings.}

\begin{figure*}[htpb]
\centering \includegraphics[width = \textwidth]
{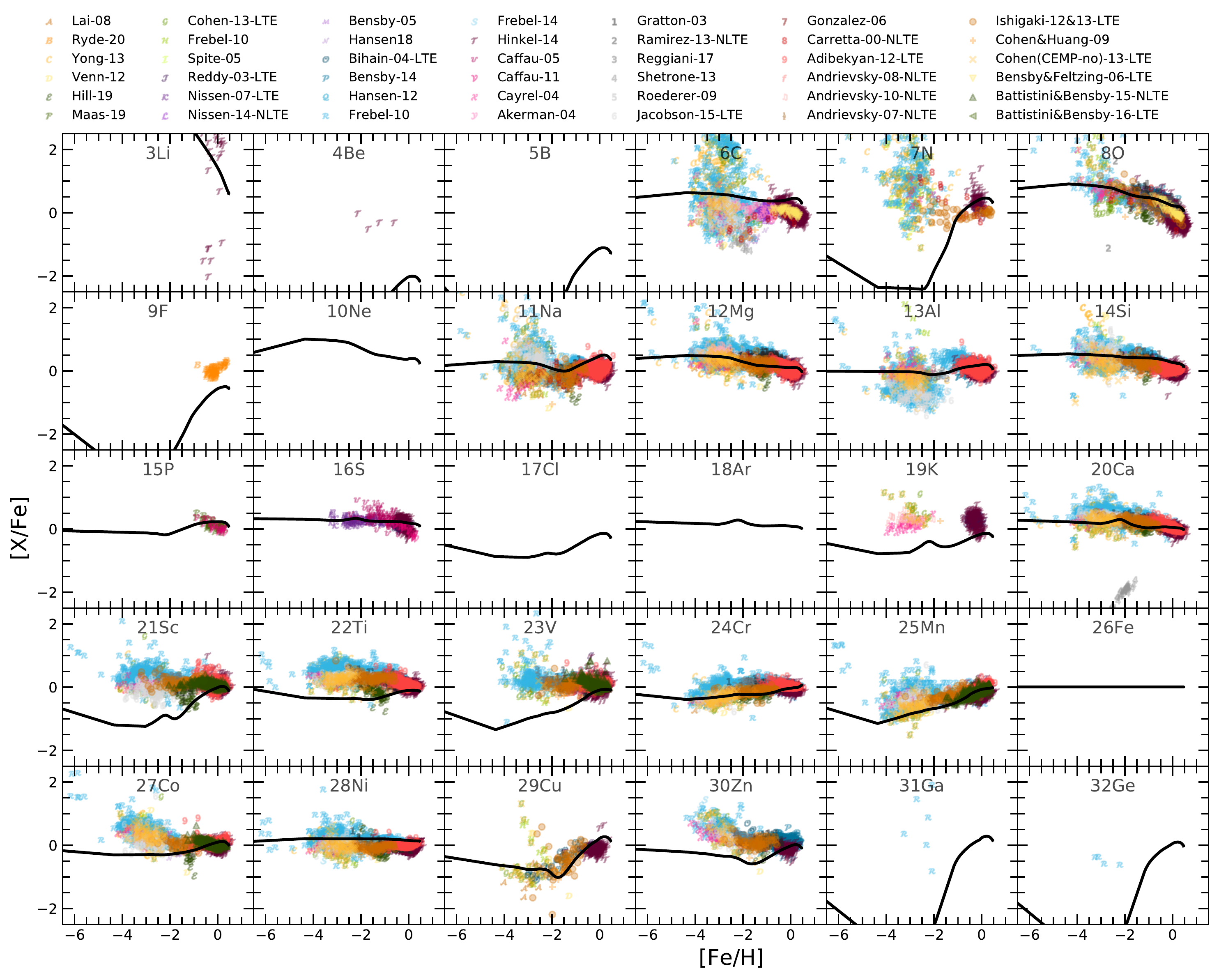}
\caption{[X/Fe]–[Fe/H] relation abundance plots for the elements up to atomic number 32. To facilitate the navigation of the table, the atomic number is shown right before the element symbol. The total abundance in the one-zone run is enriched by BBN \citep{galli13}, SNIa \citep{iwamoto99}, AGB/LIMs \citep[][with zero initial rotational velocity]{cristallo15}, and SNCC \citep[][set R, with zero initial rotational velocity]{limongi18}. The observational data scatter is labeled according to the legend on top.
}\label{fig:elemabund}
\end{figure*} 

\section{Results: {\tt GalCEM} application to a Milky Way-like One-Zone Galaxy} \label{sec:results}
{ We next present our results for a galaxy of final baryonic mass $M_{tot} = 5\times 10^{10} M_{\odot}$, that forms from an exponentially decaying gas infall with an infall timescale of 7 Gyr, no outflow, a \citet[][]{kennicutt98} SFR with $\kappa=1.$, an invariant canonical IMF \citep{kroupa01}, convolved enrichment from AGB winds and core collapse SNae with yields by \citet{cristallo11, cristallo15}, and  \citet{limongi18}, respectively, and SNIa enrichment with yields by \citet{iwamoto99} and a delay-time distribution from \citet{greggio05}, specifically the novel single-degenerate model with $\epsilon=1$.}

We first present the results from the global time-dependent quantities in \myfig \ref{fig:totphys}, namely the time-dependent evolution of total baryonic galaxy mass, total gas mass, total stellar mass, metallicity, hydrogen and helium mass on the left-hand side; meanwhile infall, SFR, and enrichment channel rates (SNCC, SNIa, and LIMs) are on the right-hand side. The evolution is tracked linearly with time expressed in Gyr. The trends of the absolute mass quantities mirrors the slope of the SFR. This is owed to the choice of linear SFR law as illustrated in Section \ref{sec:SFR}. The mass in metals is a factor of 0.05 smaller than the gas mass, which is consistent with super-solar metallicities measured in young stars.  The total stellar mass to gas mass fraction tends at the present time to a value nearly 4 times larger than the fiducial $\approx$ 10.

In \myfig \ref{fig:ageZ} we show the evolution of the metallicity and iron abundance as a linear function of time. The scatter data are taken from the stellar ages in \citet{silvaAguirre18}, the ID of the KIC stars is matched to the data from \citet{pinsonneault14} to get the iron abundance of said stars. The red lines in each plot represents a linear fit to the observational data. The black curve and blue curve are the iron abundance and metallicity, respectively. The model is in fairly good agreement with the data. It reproduces the solar abundances at the solar age. Given that the iron abundance is dominated by SNIa enrichment, we notice that [Fe/H] comes at a delay compared to the remaining metallicity (primarily oxygen) coming from SNCC and LIMs, a delay which is consistent with the SNIa rate.

\begin{figure*}[htpb]
\centering \includegraphics[width = 0.9\textwidth]
{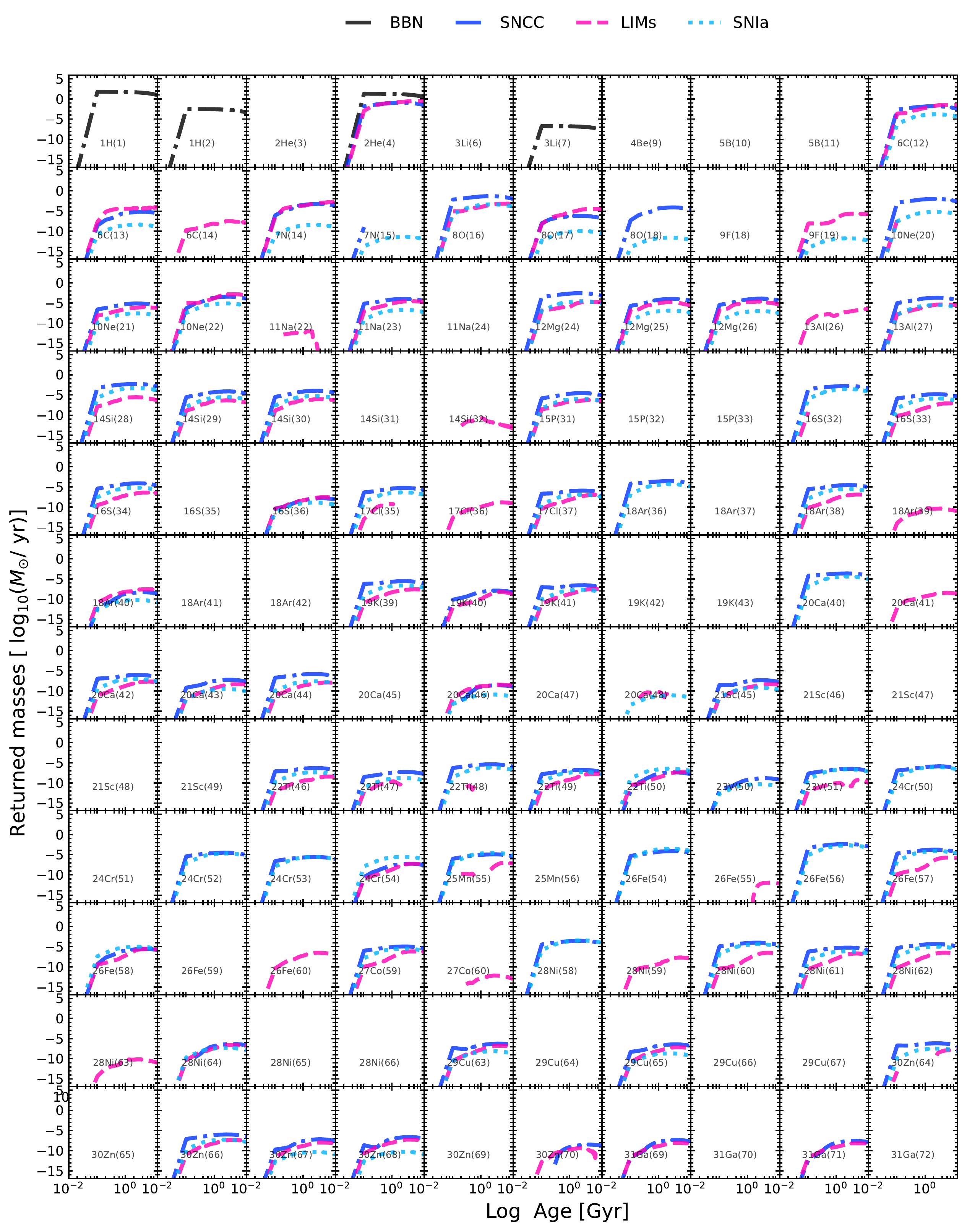}
\caption{Loglog time-dependent returned mass rates (in units of \Msun/yr) for each enrichment channel included in the one-zone run presented in this work -- which include  BBN, SNIa, LIMs, and SNCC as outlined in the Methods, Section \ref{sec:method}. The number preceding the atomic symbol represents the atomic mass.  To improve readability, we only show the first 120 isotopes, even though the full simulation computes 451 species. This cut includes every zinc nuclide (atomic number of 30), consistent with \myfig \ref{fig:elemabund}.  The full table will be shown in Gjergo et al. (2022b) in prep., with the inclusion of r-process enrichment channels.
} \label{fig:iso_rate}
\end{figure*}

\myfig \ref{fig:elemabund} is the central result of this work. We restrict our analysis to an atomic number of 30 with zinc, because heavier elements require r-process enrichment -- to be explored in a follow-up paper. Visible in the figure are also $^{31}$Ga, and $^{32}$Ge, enriched by the s-process only. \myfig \ref{fig:elemabund} shows a summary of the evolution of the [X/Fe]-[Fe/H] relation for all the elements tracked in the current paper. 
{ The model is generally in good agreement with the data, and it is consistent with other literature results constructed on similar one-zone modeling assumptions. Therefore we have reached our benchmark.

We stress the fact that the breath of stellar abundance patterns observed in the Milky Way, as well as in other galaxies, spans a rich breath of compositions and histories that cannot be reproduced by a one-zone model alone.  
We also point out at the fact that the [X/Fe]-[Fe/H] relation is only a proxy for metallicity evolution in the Galactic disk, and that the relation between iron content and metallicity breaks down for the metal poor stars found e.g. in the halo \citep[e.g., ][]{matteucci12, prantzos18}. No one-zone model is able, without treatments on dynamics or multiple infall episodes, to reproduce the average patterns for all elements. Nonetheless, the one-zone approach to GCE offers precious and reliable constraints when multiple abundance patterns are considered simultaneously -- chiefly a congruent comparison of the different timescales and rates associated with each enrichment channel, and how these inform the star formation history in a galaxy.

We briefly explain the observed patterns in \myfig \ref{fig:elemabund}, and we defer an in-depth analysis for each element to future works. Intermediate-mass elements are commonly grouped as follows: the CNO nuclei, the $\alpha$ elements, the odd-Z elements, and the iron-peak elements. Among these, the $\alpha$ elements (C, O, Ne, Mg, Si, S, Ar, Ca) are the best reproduced in literature, displaying the typical [$\alpha$/Fe] plateau at the lowest metallicities, followed by a mild decrease in the disk metallicity regime \citep[e.g.,][]{chiappini01, romano10}.  This is also the case for \GalCEM. We note however that C is often omitted from the $\alpha$ elements analysis due to its flatter [Fe/H] dependence compared to the other elements in this group \citep{prantzos18}.

The dominant carbon and oxygen isotopes are $^{12}$C and $^{16}$O. They are produced both during the H burning by the CNO cycle, but they are also the most abundant species produced during the He-burning through the 3$\alpha$ process \citep{wallerstein97}. The debate is still ongoing on what is the exact breakdown among the known astrophysical sources of enrichment, but \GalCEM\, is in agreement with \citet{andrews17} and \citet{prantzos18} on the fact that massive stars produce a larger quantity of C and O compared to AGB stars. 

Nitrogen and fluorine in \myfig \ref{fig:elemabund} behave like secondary elements. Historically, this has been an issue, particularly for nitrogen, because its observational pattern more closely resembles a primary element, while yield computations used to suggest a secondary behavior (like the one seen in \myfig \ref{fig:elemabund}). The introduction of yields by low-metallicity, rapidly rotating massive stars \citep{meynet02} solves this issue neatly \citep{limongi18} and will be a subject of further investigation with our code.

The iron-peak elements proper (Cr, Mn, Co, and Ni) are predominantly synthesized by SNIa. \myfig \ref{fig:iso_rate} shows that, in fact, these are the isotopes where the SNIa returned mass approaches most closely the returned masses by the other two enrichment channels.

}

A list of the observational papers used, and the element that the observations provide, can be found in Table \ref{tab:elemobs}. The ever-expanding \GalCEM\, library allows the user to define a list of elements and get a similar printout of observational references.

\myfig \ref{fig:iso_rate} plots  the time-dependent returned mass rates (in units of [\Msun/yr]) for each enrichment channel included in the one-zone run presented in this work. They consist of  BBN, SNIa, LIMs, and SNCC as outlined in Section \ref{sec:method}. To improve readability we only show the first 120 isotopes, even though the full simulation computes 451 species. This cut includes every zinc nuclide (atomic number of 30); i.e., we include all the elements of interest from \myfig \ref{fig:elemabund}.  Within the first Gyr, all the returned rates normalize to a given rate, pointing to the fact that the most variation in chemical evolution models occurs within the first few hundred million years.

\section{Discussion} \label{sec:discussion}

The present article has presented the features and rationale behind a new modular publicly available GCE code that computes using efficient numerical methods the convolved  integrodifferential equations of chemical evolution.

A \GalCEM\, hallmark is that it computes the enrichment of the full set of individual isotopes from multiple enrichment channels. It is sufficient to define a list with the processes one wishes to include in the simulation. 
By simply defining flags in the input class, it is possible to switch yield tabulations from a preprocessed set. The user may run a script to preprocess custom yields, or they may choose from a library of popular yields already analyzed by the \GalCEM\, team.

\GalCEM\, is suitable for studies that involve the simultaneous analysis of multiple (or all) isotopes and elements in given runs, because the code automatically generates the list of unique isotopes included in the yield tabulations. \GalCEM\, contains a  preprocessing interpolation tool that generates a multidimensional interpolation to the input yield tables for each isotope. The number of dimensions in the present work is limited to 3, namely yield, mass, and metallicity -- but the tool can accommodate extra dimensions such as initial rotational velocity. The interpolation tool may also be adapted to handle stellar lifetime tabulations.

\begin{sidewaystable*}
\vspace{-9cm}
\setlength\tabcolsep{3.5pt} 
\footnotesize  
\centering
   \caption{List of observations included in \myfig \ref{fig:elemabund}. The element investigated in each paper is marked with an $\times$. The processed public data is available for download at \url{https://github.com/egjergo/GalCEM/tree/main/galcem/input/observations/abund}}
\resizebox{\textwidth}{!}{\begin{tabular}{l*{32}{c}}
\toprule
&Li&Be&B&C&N&O&F&Ne&Na&Mg&Al&Si&P&S&Cl&Ar&K&Ca&Sc&Ti&V&Cr&Mn&Fe&Co&Ni&Cu&Zn& \\ \midrule
\citet{adibekyan12} & $\bigcirc$ & $\bigcirc$ & $\bigcirc$ & $\bigcirc$ & $\bigcirc$ & $\bigcirc$ & $\bigcirc$ & $\bigcirc$ & $\times$ & $\times$ & $\times$ & $\times$ & $\bigcirc$ & $\bigcirc$ & $\bigcirc$ & $\bigcirc$ & $\bigcirc$ & $\times$ & $\times$ & $\times$ & $\times$ & $\times$ & $\times$ & $\times$ & $\times$ & $\times$ & $\bigcirc$ & $\bigcirc$ & \\
\citet{akerman04} & $\bigcirc$ & $\bigcirc$ & $\bigcirc$ & $\times$ & $\bigcirc$ & $\times$ & $\bigcirc$ & $\bigcirc$ & $\bigcirc$ & $\bigcirc$ & $\bigcirc$ & $\bigcirc$ & $\bigcirc$ & $\bigcirc$ & $\bigcirc$ & $\bigcirc$ & $\bigcirc$ & $\bigcirc$ & $\bigcirc$ & $\bigcirc$ & $\bigcirc$ & $\bigcirc$ & $\bigcirc$ & $\times$ & $\bigcirc$ & $\bigcirc$ & $\bigcirc$ & $\bigcirc$ & \\
\citet{andrievsky07} & $\bigcirc$ & $\bigcirc$ & $\bigcirc$ & $\bigcirc$ & $\bigcirc$ & $\bigcirc$ & $\bigcirc$ & $\bigcirc$ & $\times$ & $\bigcirc$ & $\bigcirc$ & $\bigcirc$ & $\bigcirc$ & $\bigcirc$ & $\bigcirc$ & $\bigcirc$ & $\bigcirc$ & $\bigcirc$ & $\bigcirc$ & $\bigcirc$ & $\bigcirc$ & $\bigcirc$ & $\bigcirc$ & $\times$ & $\bigcirc$ & $\bigcirc$ & $\bigcirc$ & $\bigcirc$ & \\
\citet{andrievsky08} & $\bigcirc$ & $\bigcirc$ & $\bigcirc$ & $\bigcirc$ & $\bigcirc$ & $\bigcirc$ & $\bigcirc$ & $\bigcirc$ & $\bigcirc$ & $\bigcirc$ & $\times$ & $\bigcirc$ & $\bigcirc$ & $\bigcirc$ & $\bigcirc$ & $\bigcirc$ & $\bigcirc$ & $\bigcirc$ & $\bigcirc$ & $\bigcirc$ & $\bigcirc$ & $\bigcirc$ & $\bigcirc$ & $\times$ & $\bigcirc$ & $\bigcirc$ & $\bigcirc$ & $\bigcirc$ & \\
\citet{andrievsky10} &  $\bigcirc$ & $\bigcirc$ & $\bigcirc$ & $\bigcirc$ & $\bigcirc$ & $\bigcirc$ & $\bigcirc$ & $\bigcirc$ & $\bigcirc$ & $\times$ & $\bigcirc$ & $\bigcirc$ & $\bigcirc$ & $\bigcirc$ & $\bigcirc$ & $\bigcirc$ & $\times$ & $\bigcirc$ & $\bigcirc$ & $\bigcirc$ & $\bigcirc$ & $\bigcirc$ & $\bigcirc$ & $\times$ & $\bigcirc$ & $\bigcirc$ & $\bigcirc$ & $\bigcirc$ & \\
\citet{battistini15} & $\bigcirc$ & $\bigcirc$ & $\bigcirc$ & $\bigcirc$ & $\bigcirc$ & $\bigcirc$ & $\bigcirc$ & $\bigcirc$ & $\bigcirc$ & $\bigcirc$ & $\bigcirc$ & $\bigcirc$ & $\bigcirc$ & $\bigcirc$ & $\bigcirc$ & $\bigcirc$ & $\bigcirc$ & $\bigcirc$ & $\times$ & $\bigcirc$ & $\times$ & $\bigcirc$ & $\times$ & $\times$ & $\times$ & $\bigcirc$ & $\bigcirc$ & $\bigcirc$ & \\
\citet{bensby06} & $\bigcirc$ & $\bigcirc$ & $\bigcirc$ & $\times$ & $\bigcirc$ & $\times$ & $\bigcirc$ & $\bigcirc$ & $\bigcirc$ & $\bigcirc$ & $\bigcirc$ & $\bigcirc$ & $\bigcirc$ & $\bigcirc$ & $\bigcirc$ & $\bigcirc$ & $\bigcirc$ & $\bigcirc$ & $\bigcirc$ & $\bigcirc$ & $\bigcirc$ & $\bigcirc$ & $\bigcirc$ & $\times$ & $\bigcirc$ & $\bigcirc$ & $\bigcirc$ & $\bigcirc$ & \\
\citet{bensby05} & $\bigcirc$ & $\bigcirc$ & $\bigcirc$ & $\bigcirc$ & $\bigcirc$ & $\times$ & $\bigcirc$ & $\bigcirc$ & $\times$ & $\times$ & $\times$ & $\times$ & $\bigcirc$ & $\bigcirc$ & $\bigcirc$ & $\bigcirc$ & $\bigcirc$ & $\times$ & $\bigcirc$ & $\times$ & $\bigcirc$ & $\times$ & $\bigcirc$ & $\times$ & $\bigcirc$ & $\times$ & $\bigcirc$ & $\times$ & \\
\citet{bensby14} & $\bigcirc$ & $\bigcirc$ & $\bigcirc$ & $\bigcirc$ & $\bigcirc$ & $\times$ & $\bigcirc$ & $\bigcirc$ & $\times$ & $\times$ & $\times$ & $\times$ & $\bigcirc$ & $\bigcirc$ & $\bigcirc$ & $\bigcirc$ & $\bigcirc$ & $\times$ & $\bigcirc$ & $\times$ & $\bigcirc$ & $\times$ & $\bigcirc$ & $\times$ & $\bigcirc$ & $\times$ & $\bigcirc$ & $\times$ & \\
\citet{bihain04} & $\bigcirc$ & $\bigcirc$ & $\bigcirc$ & $\bigcirc$ & $\bigcirc$ & $\bigcirc$ & $\bigcirc$ & $\bigcirc$ & $\bigcirc$ & $\bigcirc$ & $\bigcirc$ & $\bigcirc$ & $\bigcirc$ & $\bigcirc$ & $\bigcirc$ & $\bigcirc$ & $\bigcirc$ & $\bigcirc$ & $\bigcirc$ & $\bigcirc$ & $\bigcirc$ & $\bigcirc$ & $\bigcirc$ & $\times$ & $\bigcirc$ & $\bigcirc$ & $\times$ & $\times$ & \\
\citet{caffau05} & $\bigcirc$ & $\bigcirc$ & $\bigcirc$ & $\bigcirc$ & $\bigcirc$ & $\bigcirc$ & $\bigcirc$ & $\bigcirc$ & $\bigcirc$ & $\bigcirc$ & $\bigcirc$ & $\bigcirc$ & $\bigcirc$ & $\times$ & $\bigcirc$ & $\bigcirc$ & $\bigcirc$ & $\bigcirc$ & $\bigcirc$ & $\bigcirc$ & $\bigcirc$ & $\bigcirc$ & $\bigcirc$ & $\times$ & $\bigcirc$ & $\bigcirc$ & $\bigcirc$ & $\bigcirc$ & \\
\citet{caffau11P} & $\bigcirc$ & $\bigcirc$ & $\bigcirc$ & $\bigcirc$ & $\bigcirc$ & $\bigcirc$ & $\bigcirc$ & $\bigcirc$ & $\bigcirc$ & $\bigcirc$ & $\bigcirc$ & $\bigcirc$ & $\times$ & $\times$ & $\bigcirc$ & $\bigcirc$ & $\bigcirc$ & $\bigcirc$ & $\bigcirc$ & $\bigcirc$ & $\bigcirc$ & $\bigcirc$ & $\bigcirc$ & $\times$ & $\bigcirc$ & $\bigcirc$ & $\bigcirc$ & $\bigcirc$ & \\
\citet{carretta00} & $\bigcirc$ & $\bigcirc$ & $\bigcirc$ & $\times$ & $\times$ & $\times$ & $\bigcirc$ & $\bigcirc$ & $\times$ & $\times$ & $\bigcirc$ & $\bigcirc$ & $\bigcirc$ & $\bigcirc$ & $\bigcirc$ & $\bigcirc$ & $\bigcirc$ & $\bigcirc$ & $\bigcirc$ & $\bigcirc$ & $\bigcirc$ & $\bigcirc$ & $\bigcirc$ & $\times$ & $\bigcirc$ & $\bigcirc$ & $\bigcirc$ & $\bigcirc$ & \\
\citet{cayrel04} & $\bigcirc$ & $\bigcirc$ & $\bigcirc$ & $\times$ & $\times$ & $\times$ & $\bigcirc$ & $\bigcirc$ & $\times$ & $\times$ & $\times$ & $\times$ & $\bigcirc$ & $\bigcirc$ & $\bigcirc$ & $\bigcirc$ & $\times$ & $\times$ & $\times$ & $\times$ & $\bigcirc$ & $\times$ & $\times$ & $\times$ & $\times$ & $\times$ & $\bigcirc$ & $\times$ & \\
\citet{cohen09} & $\bigcirc$ & $\bigcirc$ & $\bigcirc$ & $\times$ & $\bigcirc$ & $\times$ & $\bigcirc$ & $\bigcirc$ & $\times$ & $\times$ & $\times$ & $\times$ & $\bigcirc$ & $\bigcirc$ & $\bigcirc$ & $\bigcirc$ & $\times$ & $\times$ & $\times$ & $\times$ & $\times$ & $\times$ & $\times$ & $\times$ & $\times$ & $\times$ & $\times$ & $\times$ & \\
\citet[no CEMP,][]{cohen13} & $\bigcirc$ & $\bigcirc$ & $\bigcirc$ & $\times$ & $\times$ & $\times$ & $\bigcirc$ & $\bigcirc$ & $\times$ & $\times$ & $\times$ & $\times$ & $\bigcirc$ & $\bigcirc$ & $\bigcirc$ & $\bigcirc$ & $\times$ & $\times$ & $\times$ & $\times$ & $\times$ & $\times$ & $\times$ & $\times$ & $\times$ & $\times$ & $\times$ & $\times$ & \\
\citet[CEMP,][]{cohen13} & $\bigcirc$ & $\bigcirc$ & $\bigcirc$ & $\times$ & $\times$ & $\times$ & $\bigcirc$ & $\bigcirc$ & $\times$ & $\times$ & $\times$ & $\bigcirc$ & $\bigcirc$ & $\bigcirc$ & $\bigcirc$ & $\bigcirc$ & $\times$ & $\times$ & $\times$ & $\times$ & $\times$ & $\times$ & $\times$ & $\times$ & $\times$ & $\times$ & $\times$ & $\times$ & \\
\citet{frebel10} & $\bigcirc$ & $\bigcirc$ & $\bigcirc$ & $\times$ & $\times$ & $\times$ & $\bigcirc$ & $\bigcirc$ & $\times$ & $\times$ & $\times$ & $\times$ & $\bigcirc$ & $\bigcirc$ & $\bigcirc$ & $\bigcirc$ & $\bigcirc$ & $\times$ & $\times$ & $\times$ & $\times$ & $\times$ & $\times$ & $\times$ & $\times$ & $\times$ & $\bigcirc$ & $\times$ & \\
\citet{frebel14} & $\bigcirc$ & $\bigcirc$ & $\bigcirc$ & $\times$ & $\bigcirc$ & $\bigcirc$ & $\bigcirc$ & $\bigcirc$ & $\times$ & $\times$ & $\times$ & $\times$ & $\bigcirc$ & $\bigcirc$ & $\bigcirc$ & $\bigcirc$ & $\bigcirc$ & $\times$ & $\times$ & $\times$ & $\bigcirc$ & $\times$ & $\times$ & $\times$ & $\times$ & $\times$ & $\bigcirc$ & $\times$ & \\
\citet{gratton03} & $\bigcirc$ & $\bigcirc$ & $\bigcirc$ & $\bigcirc$ & $\bigcirc$ & $\times$ & $\bigcirc$ & $\bigcirc$ & $\times$ & $\times$ & $\bigcirc$ & $\times$ & $\bigcirc$ & $\bigcirc$ & $\bigcirc$ & $\bigcirc$ & $\bigcirc$ & $\times$ & $\times$ & $\times$ & $\times$ & $\times$ & $\times$ & $\times$ & $\bigcirc$ & $\times$ & $\bigcirc$ & $\times$ & \\
\citet{hansen12} & $\bigcirc$ & $\bigcirc$ & $\bigcirc$ & $\bigcirc$ & $\bigcirc$ & $\bigcirc$ & $\bigcirc$ & $\bigcirc$ & $\bigcirc$ & $\bigcirc$ & $\bigcirc$ & $\bigcirc$ & $\bigcirc$ & $\bigcirc$ & $\bigcirc$ & $\bigcirc$ & $\bigcirc$ & $\bigcirc$ & $\bigcirc$ & $\bigcirc$ & $\bigcirc$ & $\bigcirc$ & $\bigcirc$ & $\times$ & $\bigcirc$ & $\bigcirc$ & $\bigcirc$ & $\bigcirc$ & \\
\citet{hansen18b} & $\bigcirc$ & $\bigcirc$ & $\bigcirc$ & $\times$ & $\bigcirc$ & $\bigcirc$ & $\bigcirc$ & $\bigcirc$ & $\bigcirc$ & $\bigcirc$ & $\bigcirc$ & $\bigcirc$ & $\bigcirc$ & $\bigcirc$ & $\bigcirc$ & $\bigcirc$ & $\bigcirc$ & $\times$ & $\bigcirc$ & $\bigcirc$ & $\bigcirc$ & $\bigcirc$ & $\bigcirc$ & $\times$ & $\times$ & $\bigcirc$ & $\bigcirc$ & $\bigcirc$ & \\
\citet{hill19} & $\bigcirc$ & $\bigcirc$ & $\bigcirc$ & $\bigcirc$ & $\bigcirc$ & $\times$ & $\bigcirc$ & $\bigcirc$ & $\times$ & $\times$ & $\bigcirc$ & $\times$ & $\bigcirc$ & $\bigcirc$ & $\bigcirc$ & $\bigcirc$ & $\bigcirc$ & $\times$ & $\times$ & $\times$ & $\bigcirc$ & $\times$ & $\bigcirc$ & $\times$ & $\times$ & $\times$ & $\bigcirc$ & $\times$ & \\
\citet{hinkel14} & $\times$ & $\times$ & $\bigcirc$ & $\times$ & $\times$ & $\times$ & $\bigcirc$ & $\bigcirc$ & $\times$ & $\times$ & $\times$ & $\times$ & $\times$ & $\times$ & $\bigcirc$ & $\bigcirc$ & $\times$ & $\times$ & $\times$ & $\times$ & $\times$ & $\times$ & $\times$ & $\times$ & $\times$ & $\times$ & $\times$ & $\times$ & \\
\citet[][]{ishigaki12,ishigaki13}& $\bigcirc$ & $\bigcirc$ & $\bigcirc$ & $\bigcirc$ & $\times$ & $\times$ & $\bigcirc$ & $\bigcirc$ & $\times$ & $\times$ & $\bigcirc$ & $\times$ & $\bigcirc$ & $\bigcirc$ & $\bigcirc$ & $\bigcirc$ & $\bigcirc$ & $\times$ & $\times$ & $\times$ & $\times$ & $\times$ & $\times$ & $\times$ & $\times$ & $\times$ & $\times$ & $\times$ & \\
\citet{jacobson15} & $\bigcirc$ & $\bigcirc$ & $\bigcirc$ & $\times$ & $\bigcirc$ & $\bigcirc$ & $\bigcirc$ & $\bigcirc$ & $\times$ & $\times$ & $\times$ & $\times$ & $\bigcirc$ & $\bigcirc$ & $\bigcirc$ & $\bigcirc$ & $\bigcirc$ & $\times$ & $\times$ & $\times$ & $\bigcirc$ & $\times$ & $\times$ & $\times$ & $\times$ & $\times$ & $\bigcirc$ & $\times$ & \\
\citet{lai08} & $\bigcirc$ & $\bigcirc$ & $\bigcirc$ & $\times$ & $\times$ & $\times$ & $\bigcirc$ & $\bigcirc$ & $\times$ & $\times$ & $\times$ & $\times$ & $\bigcirc$ & $\bigcirc$ & $\bigcirc$ & $\bigcirc$ & $\bigcirc$ & $\times$ & $\times$ & $\times$ & $\times$ & $\times$ & $\times$ & $\times$ & $\times$ & $\times$ & $\times$ & $\times$ & \\
\citet{maas19} & $\bigcirc$ & $\bigcirc$ & $\bigcirc$ & $\bigcirc$ & $\bigcirc$ & $\bigcirc$ & $\bigcirc$ & $\bigcirc$ & $\bigcirc$ & $\bigcirc$ & $\bigcirc$ & $\bigcirc$ & $\times$ & $\bigcirc$ & $\bigcirc$ & $\bigcirc$ & $\bigcirc$ & $\bigcirc$ & $\bigcirc$ & $\bigcirc$ & $\bigcirc$ & $\bigcirc$ & $\bigcirc$ & $\times$ & $\bigcirc$ & $\bigcirc$ & $\bigcirc$ & $\bigcirc$ & \\
\citet{nissen07} & $\bigcirc$ & $\bigcirc$ & $\bigcirc$ & $\bigcirc$ & $\bigcirc$ & $\bigcirc$ & $\bigcirc$ & $\bigcirc$ & $\bigcirc$ & $\bigcirc$ & $\bigcirc$ & $\bigcirc$ & $\bigcirc$ & $\times$ & $\bigcirc$ & $\bigcirc$ & $\bigcirc$ & $\bigcirc$ & $\bigcirc$ & $\bigcirc$ & $\bigcirc$ & $\bigcirc$ & $\bigcirc$ & $\times$ & $\bigcirc$ & $\bigcirc$ & $\bigcirc$ & $\times$ & \\
\citet{nissen14} & $\bigcirc$ & $\bigcirc$ & $\bigcirc$ & $\times$ & $\bigcirc$ & $\times$ & $\bigcirc$ & $\bigcirc$ & $\bigcirc$ & $\bigcirc$ & $\bigcirc$ & $\bigcirc$ & $\bigcirc$ & $\bigcirc$ & $\bigcirc$ & $\bigcirc$ & $\bigcirc$ & $\bigcirc$ & $\bigcirc$ & $\bigcirc$ & $\bigcirc$ & $\bigcirc$ & $\bigcirc$ & $\times$ & $\bigcirc$ & $\bigcirc$ & $\bigcirc$ & $\bigcirc$ & \\
\citet{shetrone13} & $\bigcirc$ & $\bigcirc$ & $\bigcirc$ & $\times$ & $\bigcirc$ & $\bigcirc$ & $\bigcirc$ & $\bigcirc$ & $\bigcirc$ & $\bigcirc$ & $\bigcirc$ & $\bigcirc$ & $\bigcirc$ & $\bigcirc$ & $\bigcirc$ & $\bigcirc$ & $\bigcirc$ & $\times$ & $\bigcirc$ & $\bigcirc$ & $\bigcirc$ & $\bigcirc$ & $\bigcirc$ & $\times$ & $\bigcirc$ & $\bigcirc$ & $\bigcirc$ & $\bigcirc$ & \\
\citet{spite05} & $\bigcirc$ & $\bigcirc$ & $\bigcirc$ & $\times$ & $\times$ & $\times$ & $\bigcirc$ & $\bigcirc$ & $\bigcirc$ & $\bigcirc$ & $\bigcirc$ & $\bigcirc$ & $\bigcirc$ & $\bigcirc$ & $\bigcirc$ & $\bigcirc$ & $\bigcirc$ & $\bigcirc$ & $\bigcirc$ & $\bigcirc$ & $\bigcirc$ & $\bigcirc$ & $\bigcirc$ & $\times$ & $\bigcirc$ & $\bigcirc$ & $\bigcirc$ & $\bigcirc$ & \\
\citet{ramirez13} & $\bigcirc$ & $\bigcirc$ & $\bigcirc$ & $\bigcirc$ & $\bigcirc$ & $\times$ & $\bigcirc$ & $\bigcirc$ & $\bigcirc$ & $\bigcirc$ & $\bigcirc$ & $\bigcirc$ & $\bigcirc$ & $\bigcirc$ & $\bigcirc$ & $\bigcirc$ & $\bigcirc$ & $\bigcirc$ & $\bigcirc$ & $\bigcirc$ & $\bigcirc$ & $\bigcirc$ & $\bigcirc$ & $\times$ & $\bigcirc$ & $\bigcirc$ & $\bigcirc$ & $\bigcirc$ & \\
\citet{reddy03} & $\bigcirc$ & $\bigcirc$ & $\bigcirc$ & $\times$ & $\times$ & $\times$ & $\bigcirc$ & $\bigcirc$ & $\times$ & $\times$ & $\times$ & $\times$ & $\bigcirc$ & $\times$ & $\bigcirc$ & $\bigcirc$ & $\times$ & $\times$ & $\times$ & $\times$ & $\times$ & $\times$ & $\times$ & $\times$ & $\times$ & $\times$ & $\times$ & $\times$ & \\
\citet{reggiani17} & $\bigcirc$ & $\bigcirc$ & $\bigcirc$ & $\bigcirc$ & $\bigcirc$ & $\bigcirc$ & $\bigcirc$ & $\bigcirc$ & $\times$ & $\times$ & $\times$ & $\times$ & $\bigcirc$ & $\bigcirc$ & $\bigcirc$ & $\bigcirc$ & $\bigcirc$ & $\times$ & $\times$ & $\times$ & $\times$ & $\times$ & $\times$ & $\times$ & $\times$ & $\times$ & $\bigcirc$ & $\times$ & \\
\citet{ryde20} & $\bigcirc$ & $\bigcirc$ & $\bigcirc$ & $\bigcirc$ & $\bigcirc$ & $\times$ & $\times$ & $\bigcirc$ & $\bigcirc$ & $\bigcirc$ & $\bigcirc$ & $\bigcirc$ & $\bigcirc$ & $\bigcirc$ & $\bigcirc$ & $\bigcirc$ & $\bigcirc$ & $\bigcirc$ & $\bigcirc$ & $\bigcirc$ & $\bigcirc$ & $\bigcirc$ & $\bigcirc$ & $\times$ & $\bigcirc$ & $\bigcirc$ & $\bigcirc$ & $\bigcirc$ & \\
\citet{venn12} & $\bigcirc$ & $\bigcirc$ & $\bigcirc$ & $\times$ & $\bigcirc$ & $\times$ & $\bigcirc$ & $\bigcirc$ & $\times$ & $\times$ & $\bigcirc$ & $\times$ & $\bigcirc$ & $\bigcirc$ & $\bigcirc$ & $\bigcirc$ & $\bigcirc$ & $\times$ & $\times$ & $\times$ & $\times$ & $\times$ & $\times$ & $\times$ & $\times$ & $\times$ & $\times$ & $\times$ & \\
\citet{yong13} & $\bigcirc$ & $\bigcirc$ & $\bigcirc$ & $\times$ & $\times$ & $\bigcirc$ & $\bigcirc$ & $\bigcirc$ & $\times$ & $\times$ & $\times$ & $\times$ & $\bigcirc$ & $\bigcirc$ & $\bigcirc$ & $\bigcirc$ & $\bigcirc$ & $\times$ & $\times$ & $\times$ & $\bigcirc$ & $\times$ & $\times$ & $\times$ & $\times$ & $\times$ & $\bigcirc$ & $\bigcirc$ & \\
 \bottomrule
\end{tabular}}
   \label{tab:elemobs}
\end{sidewaystable*}

{ \GalCEM\, allows the user to solve the full one-zone convolution integral \citep{matteucci86} for multiple channels (AGBs, SNCC, and SNIa on this first release) without resorting to popular approximations such as IMF-averaged yields or instantaneous recycling approximations.} We map the integrand quantities onto consistent array grids in order to apply the Simpson's rule and therefore solve the integrals for each isotope and each enrichment channel. The differential equation is solved with a classic fourth-order Runge-Kutta method. Parameter dependence and algorithm speeds will be analyzed in a third paper.
In \myfig \ref{fig:elemabund} we compare \GalCEM's abundance patterns with Galactic data of main-sequence stars.  Our results are consistent with the evolution of all the intermediate elements from carbon to zinc. A library of observational data will routinely be updated on \GalCEM\footnote{\url{https://github.com/egjergo/GalCEM/tree/main/galcem/input/observations/abund}}.   We will provide an analysis of heavier elements in a second paper, where we will explore r-process candidate sites.

\software{A current version of the code is available at \url{https://github.com/egjergo/GalCEM/releases}. The results of this article can be reproduced with the {\GalCEM\, 1.0.0} package release. Archives of yield processing and results compilations can be found in the organization page: \url{https://github.com/GalCEM}.}

\section*{Acknowledgements}
We are grateful for the thorough feedback provided by the reviewer, which has helped to significantly improve the quality of this article. 
We thank Sergio Cristallo for clarifications on how to use the F.R.U.I.T.Y. database. 
We thank Kai Diethelm for crucial feedback early in the development of the numerical solutions in \GalCEM. We thank Steve Kuhlmann for providing helpful feedback on the article. 
E.G. designed and wrote \GalCEM\, and its contents in the GitHub repository. E.G. prepared the article, and coordinated the research. 
A.S. developed the preprocessing {\tt yield\_interpolation} tool, helped with refactoring \GalCEM, and set up the release of \GalCEM\, on the  Python Package Index (PyPI). 
A.R. provided feedback on code design, numerical methods, efficiency, and convergence tests. A.R. hosts the JupyterHub server at \url{https://galcem.space/}.
M.L. guided the early interpretation of the results, clarified relevant issues concerning stellar evolution, and shared resources on how to handle the yield gap in the 6-13\Msun\, mass range.
F.M. and E.S. provided invaluable guidance on GCE theory and history, as well as on the interpretation of the results.
M.K., T.K. and Y.Y. offered feedback on E.G.'s  preliminary tests that matured into \GalCEM, and hence influenced the early code design choices. 
J.L. gathered and cleaned the observational data on stellar abundances. 
X.F. promoted the endeavour to write \GalCEM,  and was involved with its development at every stage. 
X.F. provided the economic support that made this project possible. 
All co-authors helped with interpreting the results and giving feedback on the article.
E.G. and X.F. have been supported by the National Natural Science Foundation of China under grant (No.11922303) and the Fundamental Research Funds for the Central Universities (No.2042022kf1182). X.F. is supported by the Hubei province Natural Science Fund for  Distinguished Young Scholars (No.~2019CFA052). E.S. received funding from the European Union’s Horizon 2020 research and innovation program under SPACE-H2020 grant agreement number 101004214 (EXPLORE project). T.K. is supported by Grants-in-Aid for Scientific Research of Japan Society for the Promotion of Science (20K03958, 17K05459). E.G. acknowledges the support of the National Natural Science Foundation of China (NSFC) under grants No. 12041305, 12173016. We acknowledge the Program for Innovative Talents, Entrepreneur in Jiangsu. We acknowledge the science research grants from the China Manned Space Project with NO.CMS-CSST-2021-A08.

\bibliography{GalCEMbib}{}
\bibliographystyle{aasjournal}

\section*{Appendix} \label{sec:appendix}

\begin{lstlisting}[frame=single, breaklines=true, label=c15output,caption=Printout of the interpolant statistics for $^{16}$O on the trained C15 yield data]
GalCemInterpolant[z8.a16.irv0.O16]
                 (mass,metallicity)
train data description
            mass  metallicity      yield
count  86.000000    86.000000  86.000000
mean    3.053488     0.006049   0.006087
std     1.593863     0.006267   0.007744
min     1.300000     0.000048   0.000178
25%     1.500000     0.000790   0.000911
50%     2.500000     0.003000   0.002820
75%     4.000000     0.010000   0.008001
max     6.000000     0.020000   0.038573
    
train data 
         RMSE Abs: 1.83e-18
          MAE Abs: 1.13e-18
          Max Abs: 6.94e-18
         RMSE Rel: 2.78e-16
          MAE Rel: 2.37e-16
          Max Rel: 5.46e-16
\end{lstlisting}

\begin{lstlisting}[frame=single, label=lc18output,caption=Printout of the interpolant statistics for $^{16}$O on the trained LC18 yield data]
GalCemInterpolant[z8.a16.irv0.O16]
                 (mass,metallicity)
train data description
        mass   metallicity        yield
count   36.00    36.000000  3.600000e+01
mean    44.78     0.005027  5.726242e-01
std     34.27     0.007688  8.264590e-01
min     13.00     0.000018  4.099000e-07
25%     20.00     0.000140  2.308078e-03
50%     30.00     0.000995  2.082450e-01
75%     60.00     0.005883  8.216700e-01
max    120.00     0.018100  2.702400e+00

train data metrics
         RMSE Abs: 7.43e-18
          MAE Abs: 2.62e-18
          Max Abs: 2.78e-17
         RMSE Rel: 3.01e-16
          MAE Rel: 1.59e-16
          Max Rel: 9.61e-16
\end{lstlisting}

\end{document}